\documentclass[conference]{iacrtrans}  

\usepackage{custom}
\usepackage{subfigure}
\usepackage{enumitem}
\usepackage{graphics}  
\usepackage{tikz}  
\usepackage{fontawesome}
\makeatletter

\usepackage{tikz}

\newcommand{\mycircle}[1][black, fill=white]{\tikz[baseline=-0.5ex]\draw[#1,radius=3pt] (0,0) circle ;}%
\newcommand{\circW}{\mycircle[black, fill=white] }%
\newcommand{\circB}{\mycircle[black, fill=black] }%
\newcommand{\circG}{\mycircle[black, fill=lightgray] }%
\newcommand{\para}[1]{\smallskip\noindent\textbf{{#1~~}}}

\newcommand{\fullVersionOnly}[1]{#1}

\newcommand{\best}[1]{\textbf{#1}}

\newcommand{\callout}[1]{\noindent\textbf{#1}\xspace}

\author{Elie Bursztein\inst{1} \and Luca Invernizzi\inst{2} \and Karel Král\inst{2} \and Daniel Moghimi\inst{1} \and Jean-Michel Picod\inst{2} \and Marina Zhang\inst{1}}
\institute{
  Google, Sunnyvale, USA, \email{scaaml@google.com}
  \and
  Google, Zurich, Switzerland, \email{scaaml@google.com}
}
\authorrunning{E.Bursztein \and L.Invernizzi \and K.Král \and D.Moghimi \and J-M Picod \and M.Zhang} 

\begin{document}

\title[GPAM against Crypto Hardware using Long-Range Deep Learning]{
	Generalized Power Attacks against Crypto Hardware using Long-Range Deep Learning
}

\maketitle

\keywords{Deep Learning \and Side-Channel Analysis \and AES \and ECC}

\newcommand{\modelname}{GPAM}

\begin{abstract}
To make cryptographic processors more resilient against side-channel attacks, engineers have developed various countermeasures.
However, the effectiveness of these countermeasures is often uncertain, as it depends on the complex interplay between software and hardware.
Assessing a countermeasure's effectiveness using profiling techniques or machine learning so far requires significant expertise and effort to be adapted to new targets which makes those assessments expensive.
We argue that including cost-effective automated attacks will help chip design teams to quickly evaluate their countermeasures during the development phase, paving the way to more secure chips.

In this paper, we lay the foundations toward such automated system by proposing \modelname{}, the first deep-learning system for power side-channel analysis that generalizes across multiple cryptographic algorithms, implementations, and side-channel countermeasures without the need for manual tuning or trace preprocessing.
We demonstrate  \modelname{}'s capability by successfully attacking four hardened hardware-accelerated elliptic-curve digital-signature implementations.
We showcase \modelname{}'s ability to generalize across multiple algorithms by attacking a protected AES implementation and achieving comparable performance to state-of-the-art attacks, but without manual trace curation and within a limited budget.
We release our data and models as an open-source contribution to allow the community to independently replicate our results and build on them.

\end{abstract}

\section{Introduction}
\label{sec:introduction}
Cryptographic co--processors, which are widely used to perform security-sensitive operations, can be vulnerable to side-channel attacks.
These attacks aim to extract the secrets these chips safeguard, such as AES~\cite{kocher1999differential} or RSA keys~\cite{kocher1996timing}.
Broadly speaking, side-channel attacks recover secret data by collecting signals such as timing, power consumption~\cite{mangard2008power}, and electromagnetic emissions~\cite{quisquater2001electromagnetic} while the chip runs computations. 
These signals are later processed using statistical methods~\cite{chari2003template} or machine learning techniques~\cite{maghrebi2016breaking} to recover the targeted secret information (e.g., the AES key).
Attackers can then use the recovered secrets to bypass vital security features such as secure boot,\footnote{\url{{https://source.android.com/docs/security/features/verifiedboot}}}
remote attestation~\cite{FidoRemoteAttestation,moghimi2020tpm}, and identity protection.\footnote{\url{https://source.android.com/docs/security/features/biometric/measure}}

In recent years, \emph{side-channel attacks assisted with machine learning (SCAAMLs)}~\cite{jmelieDefcon} due to 
their superior accuracy, reduced manual work, and lesser need of domain knowledge, have started to replace profiling attacks such as Template attacks~\cite{choudary2014efficient,chari2003template}.
In particular, SCAAML attacks have proved to be effective at recovering AES keys, even when facing strong masking countermeasures~\cite{emmanuel2018study,maghrebi2016breaking,cagli2017convolutional}.

Despite their clear effectiveness against specific targets, several obstacles limit SCAAMLs broad adoption during the product development cycle
~\cite{picek2023sok}, where engineers need to be notified of potential leaks in a matter of hours or days to stick to the production schedule, including:
\begin{itemize}[noitemsep]
    \item \textbf{Lack of cross-algorithm generality}: SCAAMLs have not been reported to generalize beyond a single protected cryptographic algorithm~\cite{lerman2015machine,maghrebi2016breaking,cagli2017convolutional,emmanuel2018study,masure2023side,won2021back,hettwer2020deep,zhou2020deep,wu2020remove,perin2021influence,wouters2020revisiting,acharya2023information,wu2022choose,zaid2020methodology,perin2020strength,rioja2020similarities,zhang2020homogeneous,bhasin2020mind,lu2021pay}. There is currently no known machine learning technique that is able to attack both highly protected AES and ECC for example.

    \item \textbf{Lack of cross-implementation generality}: Current attacks rely on custom ML architectures tailored to a very specific target~\cite{perin2021influence,wouters2020revisiting,acharya2023information,wu2022choose,zaid2020methodology,perin2020strength}. Targeting a different implementation that uses different counter-measures requires to manually modify the attack.

    \item \textbf{High-expertise requirements}: So far SCAAMLs require expertise to not only modify the neural network architecture and its objectives but most of them also require expert manual pre-processing of the traces (e.g., \cite{lu2021pay}) so they can be used by the neural network.

\end{itemize}

In this paper, we provide the first step toward addressing these limitations by proposing a novel deep-learning architecture, \modelname{} (Generalized Power Analysis Model), that is able to perform fully automated power side-channel attacks against multiple protected algorithms, namely ECC and AES, countermeasures, and implementations.
\modelname{} is designed to work on raw traces straight from the oscilloscope doing away with the expensive requirement of expertly preprocessing traces before performing attacks. 
Our novel architecture, presented in \Cref{sec:scanet}, combines \emph{temporal patchification}~\cite{liu2022convnet} to process the very long traces generated by algorithms incorporating countermeasures, state-of-the-art Transformer encoder blocks~\cite{hua2022transformer} to efficiently identify long-range trace data relationships, and multi-task learning~\cite{ruder2017overview} to allow the model to attack masked implementations.

We demonstrate \modelname{} effectiveness by carrying out power analysis attacks against four protected ECDSA implementations in Section~\ref{sec:ecc}. 
These implementations counter-measures range from a simpler-to-defeat constant-time countermeasure to masking protections that are considered resistant to side channels attacks~\cite{coron1999resistance,perin2021keep,goudarzi2017lattice,roche2021side}.
Specifically, \modelname{} is able to recover the four most significant bits of the secret scalar with an accuracy between
71.86\% to 96.39\% depending on the targeted implementation.
At that level of accuracy, combining the model predictions confidence with a lattice attacks is enough to recover the full secret key~\cite{howgrave2001lattice,nguyen2002insecurity}, as demonstrated in Section~\ref{sec:ecdsa_attack}.
To the best of our knowledge, this is the first time that these highly-protected ECDSA implementations have been proven to be vulnerable to power side-channel attacks, demonstrating that \modelname{} architecture is not only general but also highly effective at attacking state-of-the-art hardware defenses.
We note that generalized models, such as \modelname{}, fulfill a different need than custom attack models. Custom attack models excel at uncovering vulnerabilities in high-value targets, but require the expertise of side-channel specialists. Instead, generalized models empower non-experts, such as implementation engineers, to evaluate the side-channel security of their designs without specialized attack knowledge. As such, they \emph{complement}, and not supersede, custom attack models.

In Section~\ref{sec:aes}, we show-case \modelname{}'s ability to generalize to multiple cryptographic algorithms without architectural changes by successfully recovering masked AES keys. 
When compared with state-of-the-art attacks that rely on manual trace pre-processing  and hand-tuning (which \modelname{} does not require), \modelname{} achieves comparable performance.

Last but not least in \Cref{sec:ecdsa_attack}, we demonstrate
that \modelname{} generalization capabilities also extend beyond white-box attacks by demonstrating its ability to recover hardware masked ECC scalar in a black-box settings.

Overall, the sum of our experimental results highlights that this new generation of generalized automated attacks is competitive with algorithm-specific state-of-the-art approaches for evaluating power leakage countermeasures.
Moreover as discussed in Section~\ref{sec:hypertuning}, the operational costs of adapting \modelname{} to a new target via automated hyper-tuning, a few hours of GPU time, is considerably lower than hiring side-channel experts. Our attack generality, speed, and cost-effectiveness move us closer to more secure chips by empowering design teams to incorporate automated countermeasure testing as part of the development process.

To allow the community to independently replicate our results and get us closer to the standardization of fully-automated side-channel leakage evaluations we open-source both our models and datasets under the Apache 2 Licence at \url{https://github.com/google/scaaml/tree/main/papers/2024/GPAM} (together with the full version of this paper).

\para{Ethics}
This research was intentionally performed on research implementations, not production ones.
Accordingly, these results do not warrant \fullVersionOnly{a coordinated} responsible disclosure.
\section{Background} \label{sec:background}
\fullVersionOnly{This section provides the key background information on cryptography, side-channel attacks, and deep learning needed to understand the paper.}

\subsection{Elliptic-curve cryptography} \label{back:ecdsa}
Elliptic-curve cryptography (ECC), which supports both key exchange and digital signatures, comprises public parameters, and a public/private key pair.
Public parameters include an elliptic curve $E$, a point $G$ on the curve, and the integer order $n$ of $G$ over~$E$.
The secret key $d$ is a random integer satisfying $1 < d < n-1$.
The public key is calculated as $Q = d\times G$ ($\times$ is the scalar multiplication operation supported by curve~$E$).
\fullVersionOnly{
As relevant to this paper, to generate a signature for a message hash $h$, ECDSA algorithm chooses a random secret $k$ such that $1 < k < n-1$, computes $(x, y) = k \times G$, $r = x \bmod n$, and $s = k^{-1}(h + r \cdot d) \bmod n$, and outputs $(r, s)$ as the signature pair.

It is critical for the private key $d$ and the per-message random secret $k$ to remain secret.
An attacker who acquires one instance of $k$ for a known signature can simply calculate the private key as $d = r^{-1}(s \cdot k - h) \bmod n$.
An attacker who can recover part of $k$ can apply lattice-based cryptanalysis to recover the private key from partial knowledge of $k$ from several signature-generation operations~\cite{howgrave2001lattice,nguyen2002insecurity}.
}

\subsection{Side-Channel Attack and Defense}
Side-channel attacks (SCA) target the execution of cryptographic algorithms~\cite{mangard2008power,kocher1996timing}. 
During the execution, certain physical signals may be generated by intermediate computations that depend on the secret data bits being processed. 
The attacker can build a distinguisher that identifies which signals are related to which secret bits. 
This is typically accomplished by attacking each segment of the key separately.
For instance, we build a distinguisher that can find the correct key byte for AES, which can be repeated for each of the 16 different key bytes of the 128-bit AES key.

There are two primary scenarios for constructing a distinguisher: 
In direct attacks, such as SPA~\cite{mangard2008power} or DPA~\cite{kocher1999differential}, the attacker attempts to retrieve the key from traces without prior modeling of the target. 
In profiling-based attacks, e.g., Template attacks~\cite{chari2003template}, the attacker first constructs a model based on previous observations of the target (or a similar one). 
We focus on white-box profiling-based attacks, which are valuable for assessing the security of an implementation against a strong, well-informed attacker.

One can use masking countermeasures to mitigate side-channel attacks by disrupting the statistical correlation between intermediate values and the physical signal, e.g., power consumption. 
To achieve this, implementations can generate a random value and combine it with secret parameters and intermediate values during computation. 
As a result, the computation is carried out using blinded secrets instead of cleartext ones.

A common protection for ECC implementations is to randomize the secret integer ($d$ or $k$) during scalar multiplication~\cite{coron1999resistance}.
For this, implementations can add a random multiple of the curve order $n$ to the private integer $k$ as $k' \rightarrow k + r \cdot n$.
Later on, when computing the scalar multiplication $k' \times G$ as in the signature generation, it results as $(k + r \cdot n) \times G = k \times G + r \cdot n \times G$.
Since $n \times G$ is equal to the point at infinity (the identity element), the expression simplifies to $k \times G$.
Randomizing the secret integer can also be achieved using the euclidean division $k = \lfloor k/r \rfloor \cdot r + (k \bmod r)$, or the secret integer can be divided into multiple random shares for extra security $k = k_1 + \ldots + k_m$. 
In this paper, we evaluate the security of ECC masking with single and double shares that are considered secure when the random share~$r$ is chosen in a way that $\|r\| \geq \|n\|/2$ (see~\cite{roche2021side,roche2020side,goudarzi2017lattice}) where $\|n\|$ stands for the bit-length of a natural number~$n$.
\fullVersionOnly{
These implementations include hardware-accelerated constant time scalar multiplication (CM0), additive masking (CM1), multiplicative masking (CM2), and a combination of the previous two (CM3).
For details, see \Cref{sec:protected_implementations_ecc}.
}
\subsection{Deep Learning} 
Throughout the paper, we assume a certain familiarity with standard deep-learning terms such as layer, activation function, and loss. Those terms are defined in widely available textbooks (e.g., \cite{Goodfellow-et-al-2016} and \cite{chollet2021deep}).

To process long traces efficiently \modelname{} borrows ideas from the recent advances in image patchification techniques which were introduced in vision transformers to efficiently process images~\cite{liu2022convnet}.

In terms of architecture, \modelname{} leverages the \emph{Transformer} architecture~\cite{vaswani2017attention} which is at the heart 
of the recent breakthroughs in  deep-learning including large language models (LLMs) such as chatGPT and Gemini.\todo{this next section could use a rewrite for the final version}
What makes the transformer architecture well-suited to side-channels attacks is its use of \emph{self-attention}~\cite{vaswani2017attention}, which enables the model to efficiently understand long-range dependencies and capture contextual information.
These abilities are key to build a generic and efficient side-channel attack model as it allows it to exploit complex data leaks that occur through interconnected relationships between distant data points.
In our work, we use a high-performance transformer block called GAU (Gated Attention Unit) which was introduced by Hua \etal~\cite{hua2022transformer}.
GAUs enhance the Transformer encoder block by replacing the vanilla attention and feed-forward network with a combined gating mechanisms that improve data representation and computation speed, leading to faster training and improved accuracy.
\fullVersionOnly{
Activation functions play a key role in model accuracy by introducing different form of non-linearity.
Different functions are better suited to different type of data and use-cases.
As suggested by~\cite{xie2020smooth}, we use swish~\cite{ramachandran2017swish}, a smooth activation function which is more robust in the presence of counter-measures.
Last but not least \modelname{} heavily relies on using multi-task learning~\cite{caruana1998multitask} to converge and generalize. 
}
\section{Threat model}
\label{sec:threat_model}
We assume the attacker has access to a clone of the targeted hardware and the tools to collect power traces following side channel attacks' standard assumptions~\cite{chari2003template}.

Our main threat model is the white-box model, even though, as illustrated in \Cref{sec:ecdsa_attack}, it is also able to perform well in a black-box setting.
Following SCA standard model, we also assume attacks are carried in two phases: the \emph{training} phase during which the attacker collects data using the cloned hardware to train their attacks and the \emph{attack} phase where the attacker is attempting to recover secret from the targeted device.

During the attack phase, the model exclusively processes the raw traces from the targeted device to recover the targeted secret without any access to the countermeasure parameters regardless of the threat model considered.
We ensure our dataset collection process is consistent with this modus operandi by using two different chips to collect our data: one chip is used for creating the training and testing data while the other one is used to collect the holdout dataset used for attack evaluations.

During training, we consider two threat models:
\begin{enumerate}
    \item \textbf{Black-box threat model}: In this threat model, the attacker has no knowledge of the deployed countermeasures, only knows the input and output of the cryptographic primitive and can only control the inputs (chosen text attack).
    
    \item \textbf{White-box threat model}: In this threat model, the attacker has full knowledge of the countermeasures used and they control all the protection parameters during the training phase. This is the main threat model used in this paper because it mimics the level of access that chip development teams or certification evaluators have.
    
\end{enumerate}

\section{Related Work} 
\label{sec:related_work}

\callout{Machine-learning (ML) side-channel attacks.}
Machine learning  has been repeatedly shown to be an effective approach to SCA.
For example, Lerman \etal~\cite{lerman2015machine} outperformed template attacks~\cite{chari2003template} in recovering keys from masked implementations of AES, leveraging classical ML algorithms such as support--vector machine (SVM). 
Maghrebi \etal\cite{maghrebi2016breaking} later applied deep--learning algorithms, including CNNs and LSTMs, to attack AES.
Bursztein \etal\cite{bursztein2019scaaml, burszteindc27} then proved the feasibility of full-trace attacks using deep learning.
These attacks tend to require a higher number of traces than classical attacks, though other researchers also adopted MLPs and CNNs to reduce the number of traces required~\cite{carbone2019deep,zaid2021efficiency,cagli2017convolutional}.

To improve upon these early results, research is still needed to overcome the following three challenges:

\callout{(1)~Trace preprocessing:} This remains a costly endeavor done by experts.
There are works that develop techniques to address this issue.
Notably, Won~\etal\cite{won2021back} developed a framework based on a multi-scale CNN to enable the integration of user-defined preprocessing phases. 
Hettwer \etal~\cite{hettwer2020deep} explored various image--classification metrics for finding points--of--interest in the signal.
Zhou and Standaert~\cite{zhou2020deep} proposed a technique based on residual networks for aligning SCA traces.
Wu and Picek~\cite{wu2020remove} used autoencoders to filter out noise added by mitigations such as clock jitter and random delays.
Transformations of one dimensional traces into two dimensional images and using established network architectures for images have been studied by \cite{hettwer_encoding_traces_as_images_9300289}.

Direct use of the whole traces remains rare when the traces are long.
Even very recent publications \cite{hajra2024estranet} suggest that using raw traces of tens of thousands of points is still not a solved problem.
To the best of our knowledge, the following are the only papers which directly target traces of at least 100k points using ML (the threshold is somewhat arbitrary since for each threshold there are papers which almost make it, e.g., \cite{EPRINT:GohJacSch20} with 65k point traces).
\cite{EPRINT:MBCCCDM20} target AES implementations automatically protected by code polymorphism (traces up to 160k points).
Lu \etal~\cite{lu2021pay}  developed an ML architecture (autoencoders and attention mechanism) acting directly on raw traces of up to 300k samples to target AES implementations from public datasets.
\cite{EPRINT:GohLauSch22} directly use traces of length up to 219k samples from the CHES 2020 contest.
Our model has improved the result of \cite{lu2021pay} on the ASCADv1 variable key dataset.
Our approach improves on prior art as it does not require trace preprocessing and can support very long traces, up to 16 million samples and up to 1 million points on a public dataset.

\callout{(2)~Generalizability:} Prior art has mainly focused on identifying the optimal network architecture for each device, implementation, and crypto algorithm~\cite{perin2021influence,wouters2020revisiting}.
This is an effective strategy to find an optimal solution for a specific attack configuration, but it is not clear how well it serves embedded engineers trying to identify leaks in a new product.
Some works are addressing this issue, by searching for the right ML architecture based on various tools such as Information Theory~\cite{acharya2023information}, Bayesian Optimization~\cite{wu2022choose}, and Gradient Visualization~\cite{zaid2020methodology}.
Pernin \etal\cite{perin2020strength} take a different approach, using an ensemble of ML models based on average class probabilities to improve generalization.
Whereas there are works that target multiple implementations (e.g., \cite{EPRINT:WCPB21}), to the best of our knowledge, no prior work has studied generalization across multiple algorithms (e.g., AES, ECC, RSA).
Our approach differs from prior art as we find a single architecture capable of generalizing across devices (with identical model weights), implementations, and algorithms (using the same tunable architecture), thus reducing training costs and heading toward a fully-automated SCA leakage evaluation for hardware certifications.

\callout{(3)~Portability:} Identical hardware devices, even when originating from the same production line, exhibit minute physical variations that result in differences in their power traces.
ML models that generalize across devices are superior, as during training, they do not need access to the devices they will eventually attack.
Prior research has looked into incorporating device-to-device variation into SCAAML training~\cite{rioja2020similarities,zhang2020homogeneous,bhasin2020mind}.
Our holdout datasets are also captured on different physical chips.

\callout{Side-channel attacks on ECC.}
Single-trace side-channel attacks on ECC aim to recover most of the secret bits in one execution of scalar multiplication.
This is ideal for signature schemes, like ECDSA, which performs scalar multiplications on a fresh integer every time.
However, these attacks~\cite{schindler2011exponent,jarvinen2017single,nascimento2018applying,heyszl2014clustering,weissbart2019one} have only been successful when scalar blinding has low entropy.
Most recently, Pernin~\etal~\cite{perin2021keep} showed  ML's effectiveness in unsupervised attacks,  recovering 90\% of the secret bits when scalar blinding is performed with 32 and 64 randomly-generated bits.
\cite{roche2020side}  leveraged ML to attack ECC key generation by collecting multiple traces from scalar multiplication for the same secret.

Prior work has also shown  lattice-based attack's  effectiveness in recovering keys from partial leakage collected on real cryptographic chips when no masking countermeasure is applied~\cite{moghimi2020tpm,jancar2020minerva,roche2021side}.
However, the effectiveness of SCAAML in assisting lattice attacks and bypassing stronger countermeasures is unknown.
Goudarzi \etal\cite{goudarzi2017lattice} combine lattice attacks with side-channel (using hypothetical SNR analysis) and show that attacks are probable when $\|r\|$ is 16, 32, 64 bits.
Based on these works, the current understanding in the community is that if $\|r\| \geq \|n\|/2$, the countermeasure is safe.
\fullVersionOnly{

}
To the best of our knowledge, our approach is the first to show that SCAAML driven lattice-attacks can recover the ECDSA key  when $\|r\| \geq \|n\|/2$ ($\|r\| \geq 128$ bits in our case) and even for higher-order masking with more than one share.

\section{\modelname{}}
\label{sec:scanet}
In this section, we present the \modelname{} (Generic Power Analysis Model) architecture and discuss how we train it.
\fullVersionOnly{
We start by discussing the training objective along with the metrics used to evaluate convergence.
Next, we detail \modelname{} model architecture.
Finally, we discuss relevant implementation 
details.
}

\subsection{Training objective and metrics}
\label{sec:tasks_and_metrics}
\callout{Training objective.}
\modelname{}'s training objective is to predict the value of a specific key byte $k_i$.
Following standard machine learning practices, we cast this problem as a classification problem where the model is tasked to produce the correct byte value out of 256 possibilities.
In practice, this translates to the model outputting softmax probabilities $P$ for every possible key candidate $c$, i.e., $P[k_i = c]$ for $c=\texttt{0x00},\ldots, \texttt{0xFF}$, each indicating the likelihood that the predicted key byte $k_i$ is equal to $c$.
We use categorical cross-entropy as our loss function.

\callout{Metrics.}
We use the following metrics to evaluate \modelname{} performance:
\begin{itemize}[nosep]
    \item \textbf{Accuracy}  is our main metric.
          It is defined as the categorical accuracy of the output, meaning the ratio of which the model predicts the correct output.
          The baseline accuracy for a random guess is 1/256 = 0.39\%.
          We will indicate it in the rest of the paper with the\ \randomroll symbol.

    \item \textbf{Rank} is the position of the correct value in the ranking of predicted byte values, sorted in descending order by probability.
          A rank of zero is assigned to the highest probability and 255 is assigned to the lowest probability, given that a byte has 256 possible values.

    \item \textbf{MaxRank} is the maximum rank over a set of model predictions for a batch of examples.
          The baseline MaxRank is 255.
          A value less than 255 implies that the key space required to bruteforce the correct value, in the worst case, has been successfully reduced, as each correct prediction is contained in a smaller range of values.
    
    \item \textbf{MeanRank} is the average rank of  predictions.
          The baseline MeanRank for a random guess is 127.5.
          A lower value implies that the key space was successfully reduced and that on average, the correct value is within a smaller range of values.

    \item \textbf{Confidence} reflects the standard margin sampling confidence \cite{bahri2022margin}  the difference between the highest and second-highest probabilities (i.e., the value \texttt{ prediction.sort(); prediction[-1] - prediction[-2]}).
          Intuitively, this metric is one of the most informative way to measure how much the model is confident that the predicted value is correct.
\end{itemize}

These metrics are well suited  to inform an evaluator about the presence of leakage.
For instance, one such indicator is the accuracy rising over the threshold given by the $3 \sigma$ rule.
However, the leakage detected by these metrics does not imply a successful attack.
To account for this, we also evaluate models as part of end-to-end attacks from trace collection to recovered key to understand their real-world performance.
For example, we attack ECDSA end-to-end in \Cref{sec:ecdsa_attack}, and AES in \Cref{sec:ascad_attack}.

Following machine learning best practices, we conduct experiments with the \modelname{} model on the test split (called validation split by some authors), then pick the best model according to the metrics discussed above, and evaluate only once on the holdout (which is collected on a different chip) by carrying the end to end attack.

\subsection{Architecture}  
\label{sec:architecture}

\begin{figure}[ht]
    \centering
    \includegraphics[width=\textwidth]{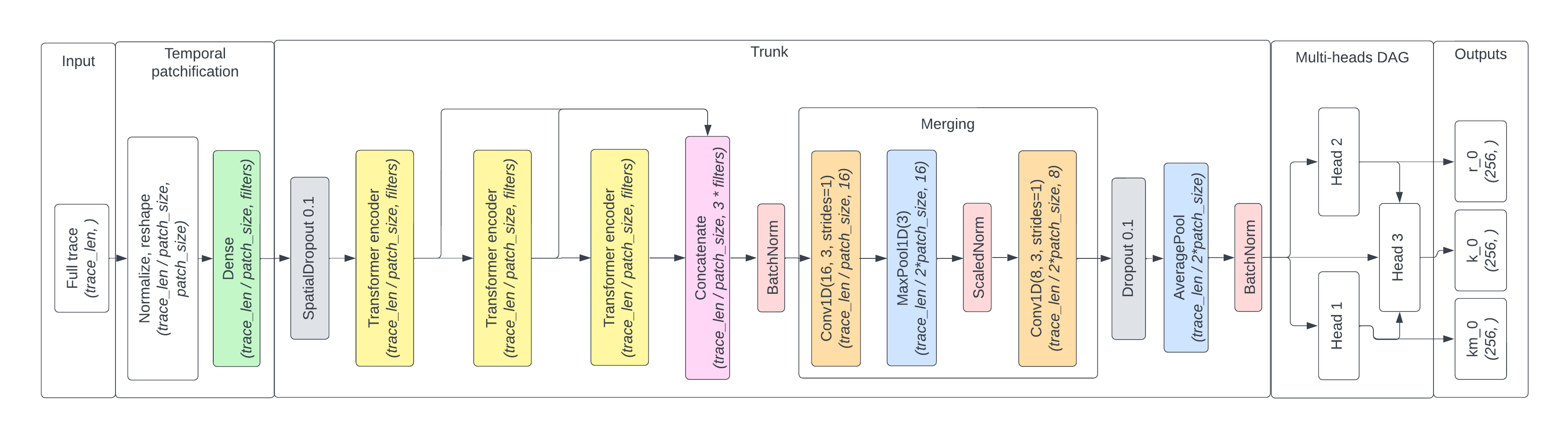}

    \caption{\modelname{} architecture for predicting $k_0$ in CM1 (see \Cref{sec:protected_implementations_ecc} for description of CM1 and the attack points $k_0, km_0, r_0$). The attacked key byte is $k_0$, with $k_0$ having related outputs $km_0$ and~$r_0$.}
    \label{fig:scanet_cm1_k_0}
    \label{fig:scanet_overview}
\end{figure}

At a high-level, \modelname{} is composed of three functional components, as depicted in \Cref{fig:scanet_cm1_k_0}: a \emph{temporal patchification} stem designed to group traces into a sequence that preserves the temporal inductive bias,
a \emph{trunk} that attends to the temporal sequence using transformer encoder blocks to extract information, and a multi-headed directed acyclic graph that is used to implement multi-task learning, i.e., predicting multiple values at once. 

\subsubsection{Temporal Patchification} 
\label{sec:temporal_patchification}

The temporal patchification stem has two goals: 
\begin{enumerate}[noitemsep,nosep]
    \item Preserve the temporal inductive bias while making the sequence faster  to process by transformer encoder blocks by grouping adjacent points -- recall here that attention computation cost is quadratic in the length of its input sequence.
    To achieve this we reshape the trace into blocks of N contiguous non-overlapping chunks, or ``patches''.
    This approach, while slightly different, is inspired by state-of-the-art image patchification techniques~\cite{liu2022convnet}.
    
    \item Potentially provide positional information to allow the transformer encoder block to perform efficiently.
    This is done by injecting global positional encoding information to the sequence~\cite{chen2021simple}.
    
\end{enumerate}

\subsubsection{Trunk} \label{sec:trunk}
The trunk's main function is to attend to the patchified sequence by extracting the latent representation of the traces needed by the heads to predict the targeted values.
To do so, \modelname{} uses a trunk made of three state-of-the-art GAU transformer encoder blocks~\cite{hua2022transformer} that are able to isolate and process long-range interacting features.
In addition to the transformer blocks, the trunk also includes a combiner module made of convolutional layers that is meant to combine the output of the three encoder blocks into a unified latent representation.
Combining the output of the encoder blocks instead of using the output of the last one, as traditionally done in NLP, is useful because each block extracts features at a different level of "abstraction". 
Those multi-level representations are commonly used in other signal processing applications such as speech recognition~\cite{chung2021w2vbert}.

\subsubsection{Heads and Relational Outputs} 
\label{sec:relational_outputs}  

The last component of \modelname{} is its multi-headed DAG (directed acyclic graph).
This component is designed to achieve two goals:
\begin{enumerate}[noitemsep,nosep]
    \item \textbf{Allow multi-task learning}: \modelname{} relies on multi-task learning~\cite{caruana1998multitask} to perform efficiently against masked implementation, as reported in Section~\ref{sec:multi_tasks_eval}.
          Multi-task learning is accomplished by not only predicting the targeted byte value but also predicting intermediate values such as the mask and random nonce values.
    \item \textbf{Inject domain expertise}: Standard multi-task learning involves jointly predicting values without establishing relation between the outputs.
          We found out, as reported in Section~\ref{sec:multi_tasks_eval}, that we can increase \modelname{} performance by representing the outputs as a DAG where intermediate outputs feed into the byte prediction output, as depicted in \Cref{fig:scanet_overview}.
          This allows the model to benefit from expert understanding and makes it easier for it to learn which intermediate values are useful for computing a given output.
          We note that defining those relations is fully configuration driven and does not require to change the model architecture or fiddle with the code.
\end{enumerate}

\subsubsection{Heads design} 
\begin{figure}[ht]
    \centering
    \vspace{-.7cm}
    \includegraphics[width=0.5\textwidth]{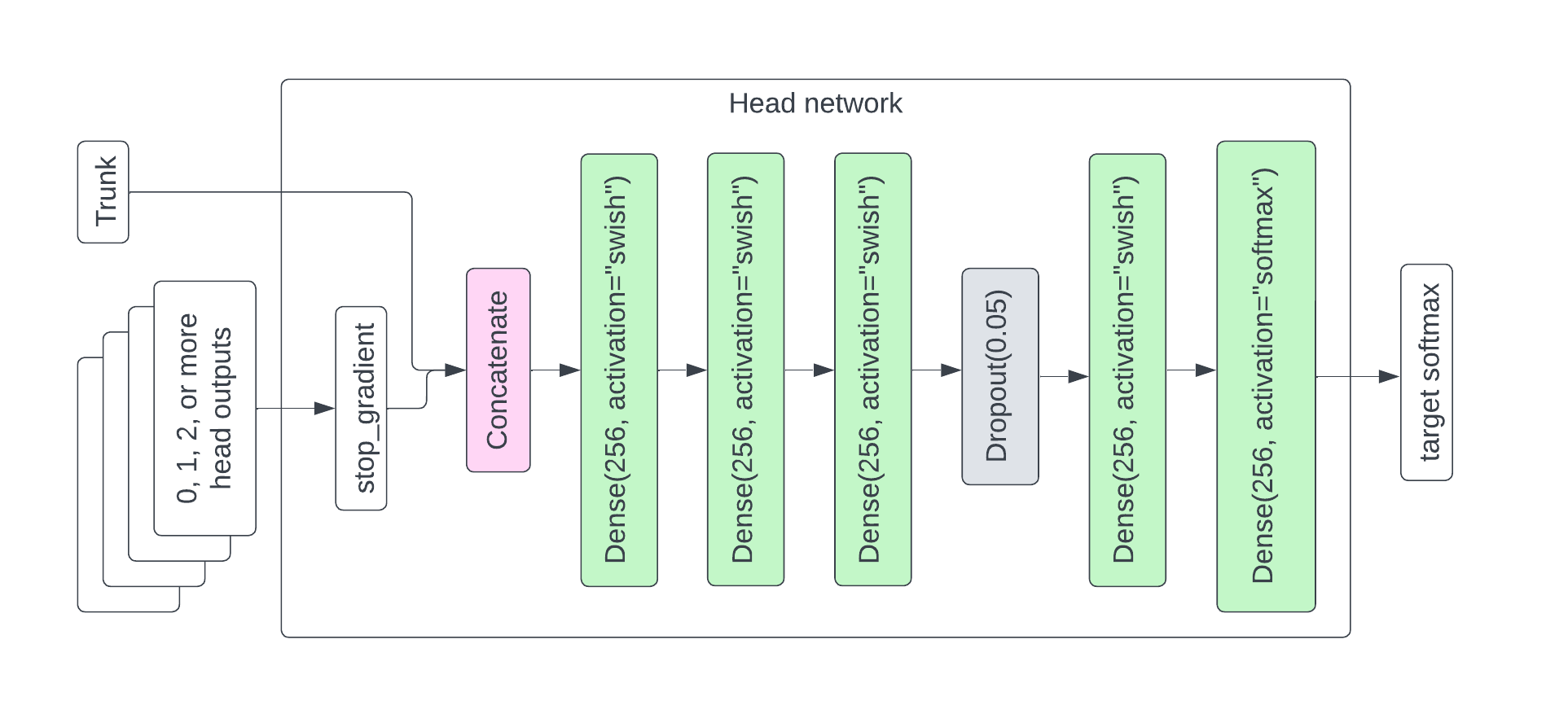}
    \caption{Single head output.}
    \label{fig:scanet_head}
    \vspace{-.5cm}
\end{figure}

Unlike standard transformer architectures, where the output is a single layer, we discovered during our architecture search that having a deeper head architecture improves model performance.
As visible in \Cref{fig:scanet_head}, \modelname{} head architecture comprises several dense layers and a single dropout layer.
Extensive initial testing during the architectural development phase revealed that adding normalization layers, residual connections, or more dropout layers did not seem to improve model performance or convergence speed.

\subsection{Hyper-parameter tuning} 
\label{sec:hypertuning}
\label{sec:model_parameters}

\modelname{} is designed to be automatically hyper-tuned (using \texttt{Keras-tuner}) to quickly adapt to new cryptographic implementations.
To minimize \modelname{} production costs, we focus on reducing the tuner search space to a minimum.
We perform a one-time extensive architecture search to isolate which parameters should be hyper-tuned and which should be considered canonical.
We emphasize that this search is a one--time cost paid by this research, and will not be run when deployed in an automated testing environment.
For example, the activation function used (Swish), the number of layers per head, and the number of GAU blocks (3) all proved to be close to optimal choices across algorithms and implementations and therefore were excluded from hyper-tuning.

\begin{table}[ht]
    \centering
\tiny
   \fullVersionOnly{\scriptsize}
    \begin{tabular}{lrrrrr}
         & \textbf{CM0} & \textbf{CM1} & \textbf{CM2} & \textbf{CM3} & \textbf{ASCADv2} \\
        \hline
        Batch size & 128 & 64 & 64 & 32 & 64 \\
        Steps per epoch & 200 & 200 & 200 & 400 & 1,000 \\
        Epochs & 25 & 500 & 500 & 500 & 150 \\
        Target learning rate & 0.0006 & 0.0006 & 0.0006 & 0.0003 & 0.00005 \\
        Merge filter 1 & 16 & 16 & 16 & 16 & 0 \\
        Merge filter 2 & 8 & 8 & 8 & 8 & 0 \\
        Trace length & 1,620,000 & 4,194,304 & 8,388,608 & 16,777,216 & 1,000,000 \\
        Patch size & 1,200 & 2,048 & 4,096 & 4,096 & 400 \\
    \end{tabular}%
    \caption{Model hyper-parameters for each targeted implementation used in this paper.}
    \label{tab:different_parameters}
\end{table}

All in all, at the end of the architecture search, \modelname{} only requires 8 parameters to be tuned for each new application.
The values of those 8 parameters for all the ECC and AES implementations targeted in this paper are reported in \Cref{tab:different_parameters}.
\begin{itemize}[nosep]
    \item \texttt{Batch size} depends purely on the hardware available for training, and should be maximized as much as the GPU memory allows (as a standard machine learning practice, doubling \texttt{batch size} implies halving \texttt{steps per epoch} and doubling \texttt{target learning rate}).
    \item \texttt{Trace length} and \texttt{patch size} depend on the size of the dataset, and the capture sample rate. The Patch size is the square root of trace length (rounded).
    \item \texttt{Target learning rate} and \texttt{number of epochs} depends on the difficulty for the optimizer to learn. Learning rate is to be searched for in logarithmic scale (good values are 0.001, 0.0001, 0.00001).
    \item \texttt{Merge filter 1} and  \texttt{Merge filter 2} are used to make the trunk output length more manageable. Merge filter 1 and 2 are searched among the powers of two, and as a rule of thumb merge filter 2 is half of the merge filter 1.
\end{itemize}

We emphase that hyper--tuning the 8 parameters is straightforward and automated.
For example, we hyper--tuned \modelname{} to attack ASCADv2 using NVIDIA RTX 4090 GPUs.
The training totaled 29 days (cumulative across GPUs), which amounts to 7 hours spent training for each parameter configuration. We note here that during hyper-tuning we do not train fully, instead we only use 10\% to 30\% of the number of epochs used for the final full training.
\fullVersionOnly{This amount of compute being enough to figure out which parameter values are the best.

}
Using a server with eight GPUs, a standard number when using NVIDIA SMX technology, this hyper-tuning search can be done under 3.6 days. For example running this training on a similar configuration on Google Cloud Compute costs under \$2.5k\footnote{This price is calculated in September 2023 with Google Cloud Pricing Calculator (\url{https://cloud.google.com/products/calculator}) running a \texttt{a2-highgpu-8g} host for 3.6 days. In reality, the GPUs offered by that configuration (A100 80GB) are slightly more performant than our 4090 cards (benchmark \url{https://lambdalabs.com/gpu-benchmarks}), so this price tag estimate is an upper bound.}.
This hypertuning can alternatively be performed on a local server with equivalent features if a local deployment is preferred, albeit for a higher startup cost.
This is a comparatively small price to pay to avoid having experts manually preprocess traces and manually tweak attack parameters, whose cumulative hourly rate can quickly surpass this dollar value.
Also, note that this hypertuning price has to be paid only once per targeted combination of hardware platform, algorithm, and countermeasure.

\callout{Weight Initialization Influence.}\label{sec:initialization_dependence}
Training instability is a well-known issue in deep learning in general, and for transformers in particular, where an unlucky initialization can lead to model collapse or sub-par performance (see for instance~\cite{nguyen2019transformers}).
Following machine-learning practices, we mitigate this issue by using a custom learning-rate scheduler.
We start from a low initial learning rate value, then steadily increase it to reach its target value, and finally decrease it with a cosine decay as the model converges~\cite{loshchilov2016sgdr}.\fullVersionOnly{
During the warm-up phase, a low learning rate allows the model to perform smaller steps along the gradient, reducing the influence of the weight initialization. 
}

\subsection{Implementation} \label{sec:implementation}
 We implement the \modelname{} architecture and conduct training using TensorFlow~\cite{tensorflow2015-whitepaper} and the Keras API~\cite{chollet2015keras}.
 We use Keras Tuner \cite{omalley2019kerastuner} to hypertune \modelname{} automatically.
 The temporal patchification code was specially designed and implemented for \modelname{}.
 The GAU layer is a custom implementation based on the pseudo-code provided in~\cite{hua2022transformer}.
 The relational output technique is our own contribution and its implementation unique to \modelname{}.
 All in all, the \modelname{} code is about 1,000 lines of Python.
 
 \begin{table}[ht]
    \centering
    \footnotesize
    \begin{tabular}{lrrrrr}
         \textbf{Dataset} & \textbf{CM0} & \textbf{CM1} & \textbf{CM2} & \textbf{CM3} & \textbf{ASCADv2} \\
         \hline
         Training time [hours] & 1 & 24 & 48 & 150 & 20 \\
    \end{tabular}
    \caption{Training time estimates.}
    \label{tab:training_times}
 \end{table}
 
 The estimated training times for attacking the various datasets used in the paper, using an NVIDIA RTX 4090 as reference GPU, are reported in \Cref{tab:training_times}.
  \fullVersionOnly{ In practice, we use multiple servers with various GPU configurations, as the total computation required to complete all our experiments presented in this paper requires over 1 year of computation. As mentioned throughout the paper, we did our best to avoid running unnecessary experiments to minimize carbon emissions.}

\section{Generic attacks against hardware-protected ECC} \label{sec:ecc}
In this section, we evaluate \modelname{}'s ability to generalize to multiple hardware-protected implementations by performing side-channel attacks against scalar multiplication on four distinct implementations of ECDSA.
\fullVersionOnly{
The targeted hardware implementation includes a constant-time implementation and three distinct algebraic masking implementations that are considered state-of-the-art protections~\cite{coron1999resistance,belaid2023high}.
We start by describing the implementations targeted, then describe our collection process. 
Next, we evaluate \modelname{}'s performance against those datasets.
Finally, we discuss how we can attack the ECDSA signature scheme by combining the partial nonce $K$ recovered via \modelname{} and a lattice attack. }

\fullVersionOnly{
Here, we answer the following questions:

\begin{enumerate}[nosep]
    \item Does \modelname{} generalize to multiple hardware implementations? (\Cref{sec:implementations_generalization_eval})
    
    \item Is Multi-task learning needed to detect leakage in a protected implementation, and if yes, which tasks are needed? (\Cref{sec:multi_tasks_eval} and \Cref{sec:multi-task_ablation})
    
    \item Are related outputs only learning a function of other outputs, or are they also using the latent representation provided by the trunk? (\Cref{sec:ablation_transformer})
    
    \item How many traces are needed for \modelname{} to successfully detect leakage in various implementations? (\Cref{sec:dataset_collection})
    
\end{enumerate}
}

\subsection{Targeted hardware implementations} 
\label{sec:protected_implementations_ecc}
Given our goal to evaluate \modelname{} against highly-protected hardware implementations, we use the NXP K82F dedicated cryptographic accelerator (LTC -- LP Trusted Crypto) as a base for all implementations to perform constant-time hardware-accelerated scalar multiplication and point addition. 
Relying on this accelerator ensures that all our implementations are not vulnerable to timing attacks or software-based leakages. 
All our ECDSA implementations use the elliptic curve FRP256\footnote{\url{https://neuromancer.sk/std/anssi/FRP256v1}} from ANSSI, but our results apply equally to other curves (e.g., NIST P-256), as the scalar multiplication algorithm is typically the same for all Weierstrass curve implementations.

\subsubsection{Countermeasures}

We evaluate the following four implementations to highlight the effectiveness of \modelname{} against increasingly stronger protections. 
Here, we use \faDesktop\ to denote computation done in software and \oncpu\ to indicate computation done on the chip. 
The $k \leftarrow \delta_L(N)$ notation indicates the generation of a random number with $N$ bits where each bit is selected independently and uniformly at random.
The four implementations considered in this section are:
\begin{enumerate}[nosep]
       \item \textbf{Constant-time execution (CM0).} 
       A simple countermeasure effective against timing attacks, but not power side channels.
       It exclusively relies on our chip's constant-time accelerated scalar multiplication, without randomizing the secret multiplier.
       
       \item  \textbf{Additive masking (CM1):} 
       This implementation is significantly more resistant to power side channel attacks compared to CM0, thanks to the addition of multiplier masking. 
       A random integer $r$ is added to $k$ so the scalar multiplication executes on the blinded secret scalar.
       More formally, it
       \begin{enumerate}[nosep]
            \item Chooses an independent 256-bit random mask $r$.
            \item  Computes the difference $km$ between the secret multiplier $k$ (the ECDSA nonce) and the mask $r$.
            \item On chip, computes $P_{km} = km \times G$ and $P_r = r \times G$.
            \item  On chip, computes $P_{km} + P_r$ and returns this value.
       \end{enumerate}
      Here is the pseudo code used to implement this scheme:
      {
\tiny
   \fullVersionOnly{\scriptsize}
        \begin{align*}
            k &\leftarrow \delta_L(256)  \tag{secret multiplier \faDesktop} \\
            r &\leftarrow \delta_L(256) \tag{random mask \faDesktop} \\
            km &= (k - r) \bmod n  \tag{$k$ masked \faDesktop} \\
            P_{km} &= km \times G \tag{scalar multiplication \oncpu} \\
            P_{r} &= r \times G \tag{scalar multiplication  \oncpu} \\
            Result &= P_{km} + P_{r} \tag{equal to $k \times G$ \oncpu}
        \end{align*}
    }%
        \item  \textbf{Multiplicative masking (CM2):} 
        In this implementation, we use a Euclidean division rather than addition computation, which is another canonical way to mask the secret scalar~\cite{belaid2023high,coron1999resistance}. 
        Formally, it:
        \begin{enumerate}[nosep]
            \item Chooses an independent 128-bit random mask $r$.
            \item Computes $k_m$, the quotient of the division of the secret multiplier $k$ and $r$. 
            It also computes the $rem$ainder.
            \item On chip, computes $P_{km} = km \times G$ and $P_r = rem \times G$.
            \item  On chip, computes $P_{km2} = r \times P_{km}$, then returns $P_r + P_{km2}$  (equal to $k \times G$).
       \end{enumerate} 
        {
   \tiny
   \fullVersionOnly{\scriptsize}
         \begin{align*}
            k &\leftarrow \delta_L(256) \tag{secret multiplier \faDesktop} \\
            r &\leftarrow \delta_L(128) \tag{random mask \faDesktop} \\
            km &= \lfloor k / r \rfloor \tag{\faDesktop}  \\
            rem &= k \bmod r \tag{\faDesktop} \\
            P_{km} &= km \times G \tag{\oncpu} \\
            P_r &= rem \times G \tag{\oncpu} \\
            P_{km2} &= r \times P_{km} \tag{\oncpu} \\
            Result &= P_r + P_{km2}  \tag{equal to $k \times G$ \oncpu}   
        \end{align*}
        }%
        \item  \textbf{Combined countermeasure (CM3):} This  combines CM1 and CM2 techniques in an attempt to further increase security with higher-order masking. 

        \begingroup
        \allowdisplaybreaks
        {
       \tiny
   \fullVersionOnly{\scriptsize}
        
        \noindent
\adjustbox{margin=-1.5cm 0 0 0 }{
\begin{minipage}[t]{.50\textwidth}
     \noindent
\begin{flalign*}
\tiny
   \fullVersionOnly{\scriptsize}
 k &\leftarrow \delta_L(256) \tag{secret multiplier \faDesktop} \\
            r_1 &\leftarrow \delta_L(256) \tag{ CM1 random mask \faDesktop} \\
            r_{2} &\leftarrow \delta_L(128) \tag{CM2 random mask \faDesktop} \\
            r_{3} &\leftarrow \delta_L(128) \tag{CM2 random mask \faDesktop } \\
            km_1 &= (k - r_1) \bmod n  \tag{\faDesktop} \\
            & \tag{CM2 (a) instead of $P_{km_1} = km_1 \times G$} \\
            km_{2} &= \lfloor km_1 / r_{2} \rfloor  \tag{\faDesktop}\\
            rem_{2} &= km_1 \bmod r_{2} \tag{\faDesktop} \\
            P_{km_{2}} &= km_{2} \times G  \tag{\oncpu} \\
            P_{km_{2}} &= r_{2} \times P_{km_{2}}  \tag{\oncpu} \\
            P_{r_{2}} &= rem_{2} \times G  \tag{\oncpu} \\
\end{flalign*}%
\end{minipage}%
\begin{minipage}[t]{.50\textwidth}
\begin{flalign*}
\tiny
   \fullVersionOnly{\scriptsize}
            P_{km_1} &= P_{r_{2}} + P_{km_{2}}  \tag{\oncpu} \\
            & \tag{CM2 (b) instead of $P_{r_1} = r_1 \times G$} \\
            km_{3} &= \lfloor r_1 / r_{3} \rfloor  \tag{\faDesktop} \\
            rem_{3} &= r_1 \bmod r_{3}  \tag{\faDesktop} \\
            P_{km_{3}} &= km_{3} \times G  \tag{\oncpu} \\
            P_{km_{3}} &= r_{3} \times P_{km_{3}}  \tag{\oncpu} \\
            P_{r_{3}} &= rem_{3} \times G  \tag{\oncpu} \\
            P_{r_1} &= P_{r_{3}} + P_{km_{3}}  \tag{\oncpu} \\
            Result &= P_{km_1} + P_{r_1}   \tag{same as $k \times G$ \oncpu} \\
            \end{flalign*}%
\end{minipage}%
   
}%

        }
        \endgroup%
\end{enumerate}%

\textbf{Random masks selection.}
We discuss the choice of the random mask $r$ (or $r_1, r_2, r_3$ in the case of CM3).
For additive masks, we choose~$r$ in a way that $\|r\| = \|n\|$, i.e., 
256 bits.
However for the multiplicative masks, we have to choose~$r$ in a way that $\|r\| < \|k\|$, otherwise, the mask would not be effective.
We choose $r$ as $\|r\| = \|n\|/2$, i.e., 128 bits, to ensure that it meets this property, but it also achieves the level of resistance to side-channel attacks, as suggested by prior work~\cite{goudarzi2017lattice,roche2021side}.
We also choose each byte of $k$ and $r$ (or $r_1, r_2, r_3$ for CM3) independently and uniformly at random for each computation to ensure we are performing attacks against implementations that do not have a bias in their random entropy.

\subsection{ Power trace collection} 
\label{sec:dataset_collection}

Our capture setup consists of a Chipwhisperer CW308 UFO board with a CW308T-K82F target board connected to it.
The firmware of the target chip was solely responsible for curve addition and scalar multiplication. 
We did not rely on any software implementation for curve arithmetic.
Scalar multiplication and point addition operations were performed by sending an integer and a point, or two points, respectively, to the LTC.
For creating labels for model training, we additionally record on the host computer the secret multiplier and random parameters used for masking each trace. 

\callout{Capture setup.}
We collect power measurements using the \emph{Teledyne LeCroy WavePro 404HD-MS}~\cite{lecroywp404hdms} oscilloscope connected to the embedded resistor shunt on the \emph{ChipWhisper NAE-CW308T-K82F}~\cite{naecw308tk82f} target board.
The oscilloscope probe is hooked to the test point TP5 of the CW308 UFO board to measure the current, while digital channel D0 is connected to the GPIO4/TRIGGER pin to get the trigger signal, which starts the capture.
We insert a 7.37MHz crystal into the X1 socket and adjust the clock source selection jumper J3 to the CRYSTAL position to provide the clock signal.
This configuration ensures that there is no correlation between the target chip clock and the oscilloscope sampling clock, resulting in asynchronous measurements.
The oscilloscope channel is set to AC coupling with a bandwidth limited to 200MHz.
\fullVersionOnly{
The first scalar multiplication is always aligned using a trigger signal for each operation, and there is no additional alignment performed. 
The trigger signal was configured to stay high during each operation performed by the LTC.
We use the first rising edge of the trigger signal as the oscilloscope trigger to start capturing.
}

\callout{Experimental leakages.}
We ensure that no UART communication is leaking by conducting one experiment where we replace captured points in the training set with Gaussian noise when the trigger signal is low.
After training a model, we observe no performance loss when compared to training on raw traces, indicating that no discernible UART leakeage is occurring in the replaced points.
\fullVersionOnly{Note that all other experiments are conducted with raw traces.
}

\callout{Datasets collected.}
\Cref{tab:list_of_datasets} provides a technical summary of the datasets generated using the implementation discussed in Section~\ref{sec:protected_implementations_ecc}.
We use the SCAAML (\cite{bursztein2019scaaml}) dataset library to store our traces and attack point values as TFRecord files. %
We ensure that no key is reused between the splits by tracking which keys were previously used.
Overall each dataset collection process takes anywhere from several weeks (CM0) to months (CM3) to complete.

Note that to ensure that the attack generalizes between chips of the same family despite potential subtle hardware variations, we use a different chip to collect the training/test splits and the holdout split.
The holdout splits, following machine learning best practices, are never used to tune the models or during experimentation.
Instead, they were reserved for the final evaluation presented in Section~\ref{sec:implementations_generalization_eval}.

Note that Table~\ref{tab:list_of_datasets} also reports the ASCADv2 dataset that we use to evaluate \modelname{} generalization across multiple algorithms in \Cref{sec:aes}.
This dataset was made public in \cite{emmanuel2018study}; we simply convert it to the SCAAML dataset format.
Since there was no apparent restriction on how to divide the samples into splits (train, test, holdout), we took a portion of consecutive samples for train, a portion for test, and a portion for holdout.

\begin{table}[htb]
  \centering\tiny
   \fullVersionOnly{\scriptsize}
    \begin{tabular}{lrrrrr}
         \textbf{Name} & \textbf{Trace } & \textbf{Train} & \textbf{Test} & \textbf{Holdout} & \textbf{File size}\\
           \textbf{} & \textbf{ length} & \textbf{length} & \textbf{length} & \textbf{length} & \textbf{ [TB]}\\
            \hline
         ECC CM0 & 1,6M & 57,344 & 8,192 & 8,192 & 0.2\\
         ECC CM1 & 5M & 194,544 & 8,192 & 8,192 & 1.5\\
         ECC CM2 & 10M & 122,880 & 8,192 & 8,192 & 2.1\\
         ECC CM3 & 17,5M & 122,880 & 8,192 & 8,192 & 3.7\\
         ASCADv2 & 1M & 640,000 & 80,000 & 80,000 & 0.9\\
    \end{tabular}
    \caption{List of ECC evaluation datasets used in this study to evaluate \modelname{} generalization to multiple hardware implementations. The names refer to protected implementations described in~\Cref{sec:protected_implementations_ecc}. The table also includes the ASCADv2 dataset collected in ~\cite{emmanuel2018study} that is used in \Cref{sec:aes} to evaluate \modelname{} generality across multiple algorithms.}
    \label{tab:list_of_datasets}
\end{table}

\subsection{Generalization over multiple implementations }
\label{sec:implementations_generalization_eval}

Overall, \modelname{} can successfully attack all four ECC hardware implementations in white-box settings using multi-task relational outputs training as reported in Table~\ref{tab:best_results_ecc_whitebox}.
These results are computed on the holdout splits which, as discussed previously, are captured on a different chip of the same family and were not used for model tuning or any other experiments discussed later in this section.
As discussed in the threat model section (\Cref{sec:threat_model}), white-box setting means that the model had access to the intermediate values (masks and random values) during training.

Note that we only evaluate \modelname{} on the initial (most-significant byte $k_{0}$), middle  ($k_{15}$), and last (least-significant byte $k_{31}$) byte of each implementation, as those bytes are representative of \modelname{} performance against those implementations. We make this choice because these experiments are resource intensive, and the goal of \modelname{} is to identify leakage.

\begin{table}[htb]
  \centering\tiny
   \fullVersionOnly{\scriptsize}
    \begin{tabular}{lrrrr}
         \textbf{Dataset} & \textbf{Attack } & \textbf{Accuracy} & \textbf{MeanRank} & \textbf{MaxRank}\\
                  \textbf{} & \textbf{ point} & \textbf{ [\%]} & \textbf{} & \textbf{}\\
        \hline
         CM0 & $k_0$ & 100.00 & 0 & 0 \\
         CM0 & $k_{15}$ & 100.00 & 0 & 0 \\
         CM0 & $k_{31}$ & 100.00 & 0 & 0 \\
         \midrule
         CM1 & $k_0$ & 78.80 & 0.75 & 192 \\
         CM1 & $k_{15}$ & 93.20 & 0.31 & 253 \\
         CM1 & $k_{31}$ & 92.98 & 0.24 & 218 \\
         \midrule
         CM2 & $k_0$ & 66.22 & 1.40 & 254 \\
         CM2 & $k_{15}$ & 0.30 \randomroll & 127.29 & 255 \\
         CM2 & $k_{31}$ & 11.31 & 8.76 & 233 \\
         \midrule
         CM3 & $k_0$ & 8.60 & 19.77 & 255 \\ 
         CM3 & $k_{15}$ & - & - & - \\
         CM3 & $k_{31}$ & 0.37 \randomroll & 127.89 & 255 \\
    \end{tabular}
    \caption{\modelname{} white-box key byte recovery success rate on the four ECC hardware protected implementations holdout splits.}
    \label{tab:best_results_ecc_whitebox}
\end{table}

As expected, as the strength of the protection increases, the model accuracy decreases to the point where for CM3, only the initial byte can be attacked successfully. 
Note that an accuracy of $0.4\%$ is close to random chance. We hypothesize that increasing \modelname{}'s performance against stronger countermeasures requires more training data, not increased model capacity.
This is empirically supported by our experiment in Section~\ref{sec:num_traces_eval}, which looks at model accuracy as a function of the number of training traces, showing that \modelname{}'s accuracy against CM3 only starts to rise past 100k traces.
Furthermore, looking at the MaxRank results, it is clear that the model did not fully generalize for CM1, CM2 and CM3, as it is close to its 255 upper bound.

\begin{table}[tb]
    \centering
   \tiny
   \fullVersionOnly{\scriptsize}
    \begin{tabular}{lrrrr}
         \textbf{Dataset} & \textbf{Attack } & \textbf{Accuracy} & \textbf{MeanRank} & \textbf{MaxRank}\\
                  \textbf{} & \textbf{ point} & \textbf{ [\%]} & \textbf{} & \textbf{}\\
         \hline
         CM0 & $k_0$ & 100 & 0 & 0 \\

         CM1 & $k_0$ & 0.29~\randomroll & 127.5 & 255 \\

         CM2 & $k_0$ & 22.81 & 5.55 & 231 \\
    \end{tabular}
    \caption{\modelname{} black-box key byte recovery success rate on the first three ECC hardware protected implementations using the holdout splits.}
    \label{tab:best_results_ecc_blackbox}
\end{table}%

In Table~\ref{tab:best_results_ecc_blackbox}, we report \modelname{} results under black-box settings on the holdout splits.
Unlike when using the white-box settings, \modelname{} is not always able to successfully recover even the initial byte $k_0$.
Interestingly, \modelname{} fails to recover CM1 $k_0$ but is able to recover CM2 $k_0$, which is surprising given that in the white-box setting, CM1 appears to be easier than CM2.
We hypothesize that CM1 is harder in the black-box setting because its random mask uses 256 bits whereas CM2 only uses 128 bits.
Having access to intermediate values seems to make the size of the random mask irrelevant in the white box setting but a strong defense in the black box setting.

\subsection{Multi-task effectiveness evaluation}
\label{sec:multi_tasks_eval}

In the following set of experiments, we study whether using multi-task learning improves \modelname{}'s accuracy.
In particular, we are interested in understanding which additional tasks beyond the key byte prediction improve model accuracy, if any.
To not taint our holdout splits, the results reported in this section are computed on the test splits. 
Once again we only perform experiments on representative bytes to limit the computation time, namely the initial bytes ($k_0$, $k_1$, $k_2$), middle one ($k_{15}$) and final ones ($k_{29}$, $k_{30}$, $k_{31}$).
For that same reason, we also only target the middle of the road dataset CM2 and reserve the study of CM1 for the ablation study discussed later in the paper in Section~\ref{sec:multi-task_ablation}.

Overall, there are two types of additional tasks that can be included in the training:
\emph{Adjacency predictions} and \emph{Intermediate predictions}.
\emph{Adjacency predictions} ask the model to predict the key bytes on the left and right of  targeted bytes with the hope it helps with carry issues and generalization. %
\emph{Intermediate predictions} involve the model predicting the value of intermediate computation points, including mask and random nonces values.

The model can operate in two modes when performing multi-task learning using intermediate values: the \emph{multi-outputs} mode and \emph{relational outputs} mode.
In the \emph{multi-outputs} mode, the model outputs all the asked values without any interaction between outputs.
This is the classical form of multi-task learning used by many models to boost generality and accuracy. 
In the \emph{relational output} mode, as illustrated in \Cref{fig:scanet_cm1_k_0}, we create a directed acyclic graph between the heads to model expert knowledge of how the outputs relate to each other according to the protecting algorithm (CM1-CM3).
Obviously this type of knowledge is only available in white box testing conditions.

\callout{Notation.}
To make the results tables easier to understand, we are using the following visual convention to distinguish between the various relation conditions:
\begin{itemize}[noitemsep,nosep]

\item Circles are byte indexes centered at the column index (if the column byte index is $i$ then the circles represent $[i-1, i, i+1]$).

\item A white circle \circW at position $j$ means not leveraging multi-task learning.

\item A gray circle \circG at position $j$ means using multi-task learning.

\item A black circle \circB means using relational outputs learning.

\end{itemize}

\fullVersionOnly{
Here are a few examples of such notations for the column $k_2$ of \Cref{tab:adjacent_intermediates_acc}:

\begin{itemize}[noitemsep,nosep]
    \item[\circW\circW\circW] Outputs: $k_2$, relations: $[]$. Byte $k_2$ is targeted and no multi-task learning is used.
    
    \item[\circW\circG\circW] Outputs: $k_2, km_2, r_2$, relations $[]$. The model operates in the multi-task learning mode and predicts the values of $k_2, km_2,$ and $r_2$. 
    
    \item[\circW\circB\circW] Outputs: $k_2, km_2, r_2$, relations $[km_2 \rightarrow 
    k_2, r_2 \rightarrow k_2]$. The model uses relational outputs to predict the values of $k_2, km_2,$ and $r_2$. The values of $km_2$ and $r_2$ are fed into $k_2$.
    
    \item[\circB\circB\circW] Outputs: $k_2, km_1, r_1, km_2, r_2$, relations $[km_1 \rightarrow k_2, r_1 \rightarrow k_2, km_2 \rightarrow k_2, r_2 \rightarrow k_2]$. 
    The model uses relational outputs and adjacency relations to predict the values of $k_2, km_2,$ and~$r_2$. The values of $km_1$, $r_1$, $km_2$ and~$r_2$ are fed into $k_2$.
\end{itemize}
}

\callout{Results.}
Overall, we observe that multi-task learning is needed for the attack to succeed as reported on CM1 in Table~\ref{tab:adjacent_intermediates_acc}.
Without any relations, denoted using the \circW symbol, the model predictions are unable to exceed random chance ($k_0$, $k_1$, $k_2$, $k_{15}$, $k_{29}$) or barely exceed it ($k_{30}$ and $k_{31}$).
Using the simplest form of multi-task learning, denoted using the \circG symbol, the model obtains high accuracy for almost all the bytes except $k_{15}$.
Using relational-output, denoted using the \circB symbol, the model accuracy improves further overall compared to using multi-task learning.

\begin{table}[htb]
  \centering\tiny
   \fullVersionOnly{\scriptsize}
    \begin{tabular}{lrrrrrrr}
         \textbf{Relations} & \textbf{$k_0$} & \textbf{$k_1$} & \textbf{$k_2$} & \textbf{$k_{15}$} & \textbf{$k_{29}$} & \textbf{$k_{30}$} & \textbf{$k_{31}$} \\
         \midrule
         \circW\circW\circW & 0.39 & 0.39 & 0.20 & 0.20 & 0.39 & 0.78 & 0.59 \\
         \circW\circG\circW & 81.50 & \best{79.20} & 85.45 & 43.36 & \best{86.62} & 86.13 & 92.68 \\
         \circW\circB\circW & 85.35 & 71.88 & \best{88.09} & \best{85.64} & 85.64 & 87.21 & 92.58 \\
         \circB\circB\circW & -- & 67.68 & 81.45 & 40.82 & 82.52 & 83.38 & \best{93.65} \\
         \circW\circB\circB & \best{87.40} & 74.80 & 83.98 & 63.67 & 82.03 & 87.70 & -- \\
         \circB\circB\circB & -- & 63.38 & 80.96 & 41.11 & 84.08 & \best{88.48} & -- \\\bottomrule
    \end{tabular}
    \caption{\modelname{} accuracy in \% on CM1 test dataset when trained using various forms of multi-task learning.}
    \label{tab:adjacent_intermediates_acc}
\end{table}

Conversely, the effectiveness of adjacency relations is marginal.
As reported in Table~\ref{tab:adjacent_intermediates_acc}, adjacency relations introduce model instability with the accuracy slightly increasing on some bytes (e.g., $k_0$, $k_{30}$, $k_{31}$) but decreasing significantly on others (e.g., $k_{15}$).
Given  our goal to have a stable fully automated attack, we decided against using adjacency relations.
\fullVersionOnly{
We leave it as future work to make better use of them.
}

\callout{Effect of multi-task learning on model convergence.}
\label{sec:convergence}
The positive impact of multi-task learning is best visualized by looking at how each of the output accuracies improve as training progresses.
Regardless of the hardware implementation, we observe that outputs start to converge one after the other. 
\begin{figure}[htb]
    \centering
    \subfigure[CM1]{\label{fig:ecc_cm1_val_acc_k_0}
    \vspace{-1.8cm}\includegraphics[width=0.31\textwidth]{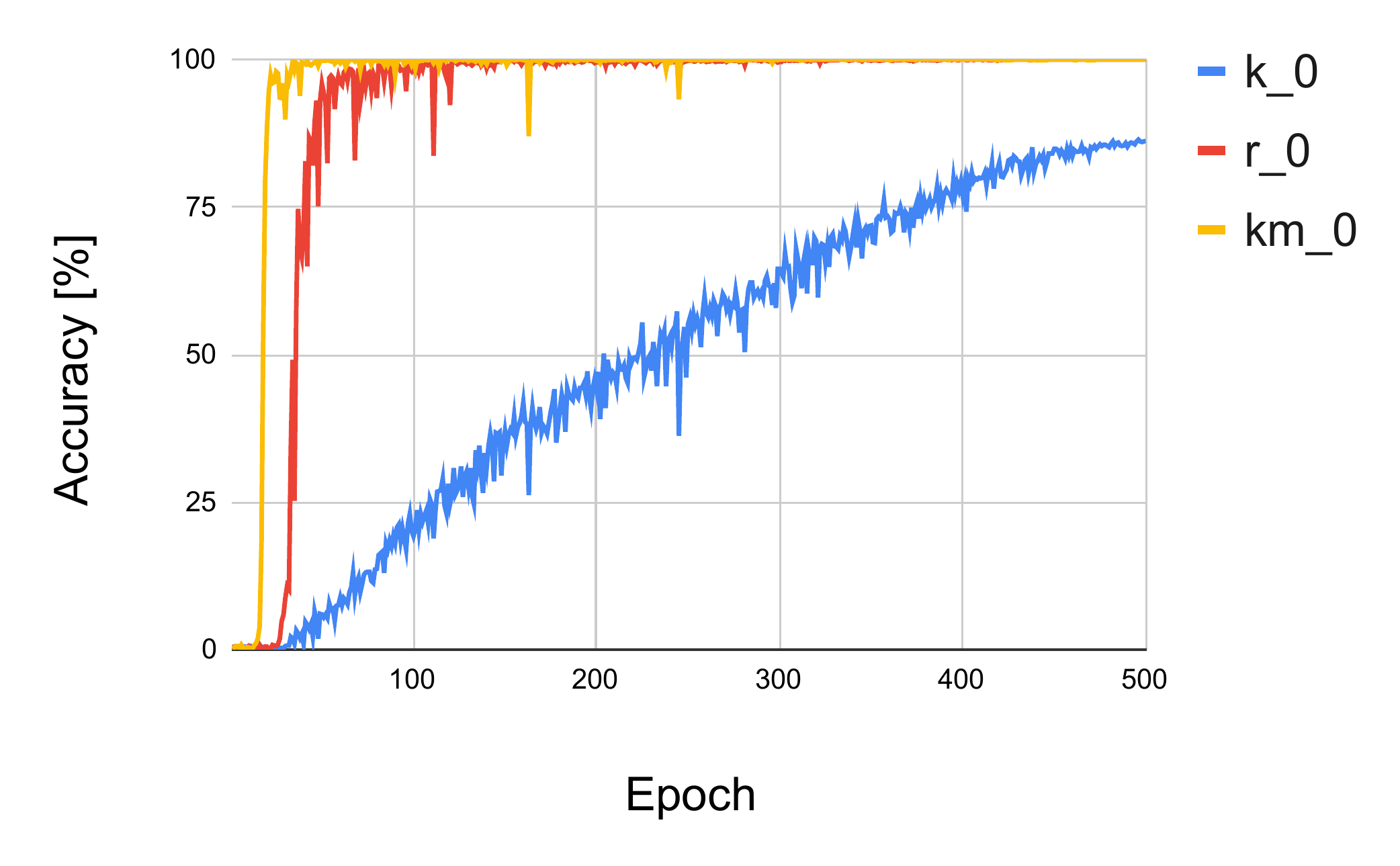}}%
    \subfigure[CM2]{\label{fig:ecc_cm2_val_acc_k_0} \vspace{-1.8cm}\includegraphics[width=0.31\textwidth]{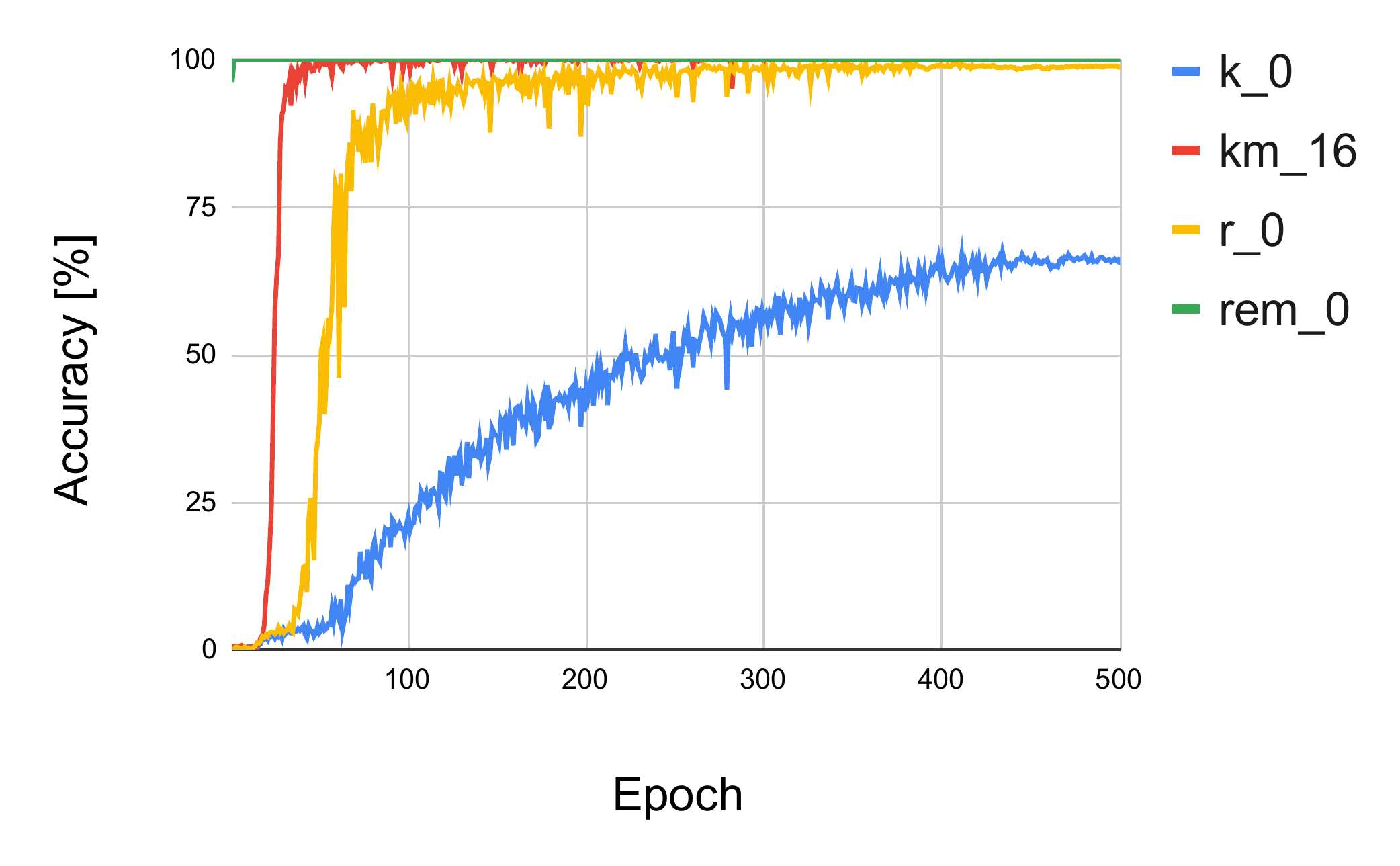}}%
    \subfigure[CM3]{\label{fig:ecc_cm3_val_acc_k_0} \vspace{-1.8cm}\includegraphics[width=0.31\textwidth]{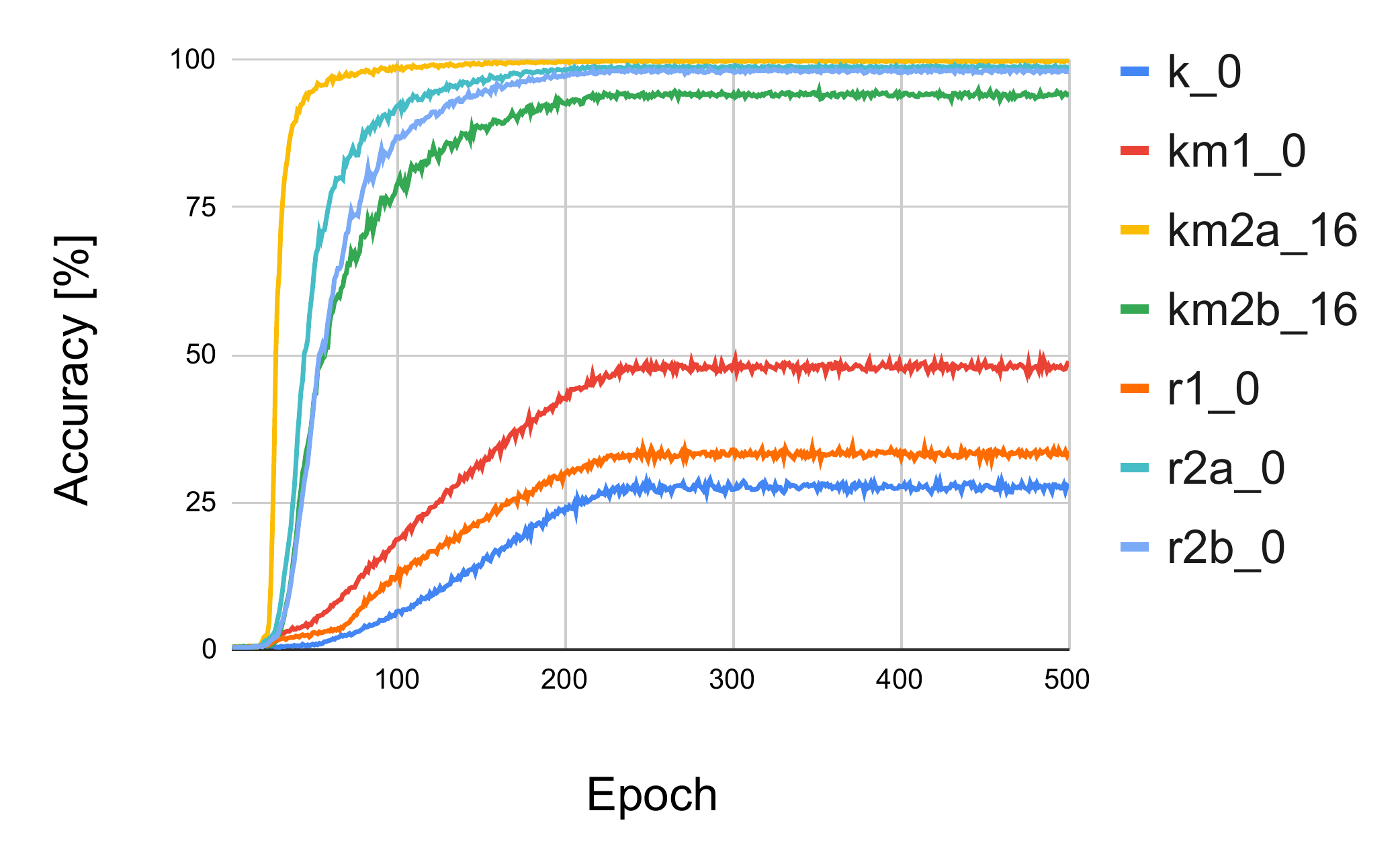}}
    \caption{ECC validation accuracy for $k_0$ when using relational outputs.}
\end{figure}
For example, as visible in Figure~\ref{fig:ecc_cm1_val_acc_k_0}, we observe that the mask prediction ($km_0$) accuracy rises before the random nonce ($r_0$) prediction accuracy improves, and the key value ($k_0$) prediction starts converging only after both the mask and the random nonce have reached a high accuracy. 
The same effect is observed for CM2 (\Cref{fig:ecc_cm2_val_acc_k_0}) and CM3 (\Cref{fig:ecc_cm3_val_acc_k_0}).

Additionally, we observe that in each case the model first learns to predict the mask values and then the random nonces. 
This behavior is consistent with the hypothesis that multi-task learning is critical to create generalized SCAAML attacks as it allows models to learn to "unpack" protections one step at a time.
It also supports the hypothesis that black-box attacks are significantly harder, because models greatly benefit from the extra information. 
Last but not least, this behavior seems to confirm the effectiveness of higher-order masks against advanced side channel attacks such as SCAAML.

\subsection{Multi-task ablation study}
\label{sec:multi-task_ablation}
In this section, we perform an ablation study to better understand which intermediate values are needed for the attacks to succeed on CM1 and CM2.
We exclude CM0, as there are no intermediate values.
We also exclude CM3, as \modelname{}'s relatively low accuracy on this dataset makes it hard to confidently separate the results, and CM3 experiments would take roughly 16 months of computation.

\begin{table}[htb]
  \tiny
   \fullVersionOnly{\scriptsize}\centering
    \begin{tabular}{lrrrr}

        \textbf{Target} &  \textbf{Dependency} &  \textbf{Accuracy [\%]} &  \textbf{MeanRank} &  \textbf{MaxRank}\\
        \hline
        $k_0$ & $km_0, r_0$ & \best{86.26} & \best{0.17} & \best{9}\\
        $k_0$ & $km_0$ & 0.39 & 130.00 & 255\\
        $k_0$ & $r_0$ & 86.60 &  0.14 &  4\\
        $k_0$ & Nothing & 0.29 & 127.90 & 255\\
        \midrule
        $k_{15}$ & $km_{15}, r_{15}$ & \best{44.00} & \best{1.07} & \best{58}\\
        $k_{15}$ & $km_{15}$ & 0.48 & 125.60 & 255 \\
        $k_{15}$ & $r_{15}$ &  0.29 &  126.20 & 255  \\
        $k_{15}$ & Nothing & 0.48 & 125.69 & 255 \\
        \midrule
        $k_{31}$ & $km_{31}, r_{31}$ & 92.87 & 0.09 & 15\\
        $k_{31}$ & $km_{31}$ & 0.48 & 123.46 & 255\\
        $k_{31}$ & $r_{31}$ & \best{93.85} & \best{0.07} & \best{7}\\
        $k_{31}$ & Nothing & 0.20 & 126.94 & 255\\

    \end{tabular}
    \caption{CM1 relational outputs ablation.}
    \label{tab:relation_ablation_cm1}
\end{table}

For CM1, as reported in Table~\ref{tab:relation_ablation_cm1}, removing the prediction of the mask ($r_{*}$) at training time results in the model being unable to successfully attack CM1. 
Removing the prediction of the random nonce ($km_{*}$) drastically reduces the accuracy of the model for the middle key byte but has no effect on the initial and last byte.
We are not sure why this happens.

\begin{table}[htb]
    \tiny
   \fullVersionOnly{\scriptsize}\centering
    \begin{tabular}{lrrrrr}
  
        \textbf{Target} &  \textbf{Dependency} &  \textbf{Accuracy [\%]} &  \textbf{MeanRank} &  \textbf{MaxRank} \\
        \midrule
        $k_0$ & Nothing & 18.26 & 8.44 & 235 \\
        $k_0$ & Outputs: $r_0, km_{16}, rem_{16}$ & 52.44 & \best{1.78} & 235 \\
        $k_0$ & $r_0, km_{16}, rem_{16}$ & \best{66.50} & 1.99 & 252 \\
        $k_0$ & $r_0, km_{16}$ & 66.50 & 2.14 & 254 \\
        $k_0$ & $r_0, rem_{16}$ & 39.16 & 4.47 & 244 \\
        $k_0$ & $km_{16}, rem_{16}$ & 32.40 & 4.21 & 223 \\
        $k_0$ & $r_0$ & 39.26 & 4.43 & 247 \\
        $k_0$ & $km_{16}$ & 36.23 & 2.63 & \best{160} \\
        $k_0$ & $rem_{16}$ & 11.72 & 15.37 & 203  \\
        \midrule
        $k_{31}$ & Nothing & 0.39 & 125.60 & 255  \\
        $k_{31}$ & Outputs: $r_{15}, km_{31}, rem_{31}$ & 9.00 & \best{6.74} & 192  \\
        $k_{31}$ & $r_{15}, km_{31}, rem_{31}$ & 10.40 & 10.60 & 238  \\
        $k_{31}$ & $r_{15}, km_{31}$ & \best{10.74} & 9.90 & \best{142}  \\
        $k_{31}$ & $r_{15}, rem_{31}$ & 8.49 & 10.33 & 207  \\
        $k_{31}$ & $km_{31}, rem_{31}$ & 8.59 & 8.58 & 195  \\
        $k_{31}$ & $r_{15}$ & 9.27 & 8.99 & 155  \\
        $k_{31}$ & $km_{31}$ & 1.36 & 41.38 & 215  \\
        $k_{31}$ & $rem_{31}$ & 8.59 & 9.90 & 165  \\
    \end{tabular}
    \caption{CM2 relational outputs ablation.}
    \label{tab:relation_ablation_cm2}
    
\end{table}

We only target $k_0$ and $k_{31}$ in \Cref{tab:relation_ablation_cm2}, as the model is only able to target the most and least significant bytes of CM2.

\subsection{Trunk ablation study}
\label{sec:ablation_transformer}
\label{sec:ablation_of_trunk}

In this ablation study, we evaluate whether the relational outputs benefit from the trunk output.
To validate this hypothesis, we try to predict $k_{15}$ but with only using the output of the $km_{15}$ and $r_{15}$ heads not the output of the trunk.
Additionally we prevent the model from cheating and encoding $k_{15}$ prediction information in the intermediate outputs by applying a stop gradient on the $km_{15}$ and $r_{15}$ output layers.
For this ablation study, the head outputting $k_{15}$ looks like the one in \Cref{fig:scanet_head} when one removes the ``Trunk'' input.

\begin{table}[ht]
\centering\tiny
   \fullVersionOnly{\scriptsize}
    \begin{tabular}{lrrr}
        \textbf{Trunk} & \textbf{Accuracy [\%]} & \textbf{MeanRank} & \textbf{MaxRank} \\
        \hline
        yes & \best{85.64} & \best{0.52} & 198  \\
        no & 30.6 & 1.53 & \best{126}  \\
    \end{tabular}
    \caption{CM1 ablation of trunk output for the $k_{15}$ head.}
    
    \label{tab:trunk_ablation}
\end{table}

When the key byte prediction head is not directly connected to the trunk, the model exhibits an accuracy drop of 55 percentage points as reported in Table~\ref{tab:trunk_ablation}.
Those results strongly support the hypothesis that the transformer encoder blocks' latent representations  and compute power are critical for accurate prediction.

\subsection{Dataset size impact evaluation}
\label{sec:num_traces_eval}

In this section, following the insights of~\cite{kaplan2020scaling} where the authors showed that transformer models are bound by capacity, compute, or data, we attempt to determine which of these factors is limiting \modelname{} performance.
We already know, thanks to the experiment ran in Section~\ref{sec:convergence}, that \modelname{} is not bounded by compute since the performance plateaus after a few hundred epochs - see \Cref{fig:ecc_cm1_val_acc_k_0} for example.
Accordingly, to decide whether the model is capped by the amount of data available or the model capacity/architecture, we trained the same model on an increasing number of traces from CM1, CM2, and CM3.
We take 10\%--100\% of the 122,880 available examples and learn to predict $k_0$.

\begin{figure}
    \centering
    \subfigure[Accuracy]{%
        \label{fig:parts_of_ds_2rounds_k_0_acc}%
           \vspace{-.4cm}
        \includegraphics[width=0.47\textwidth]{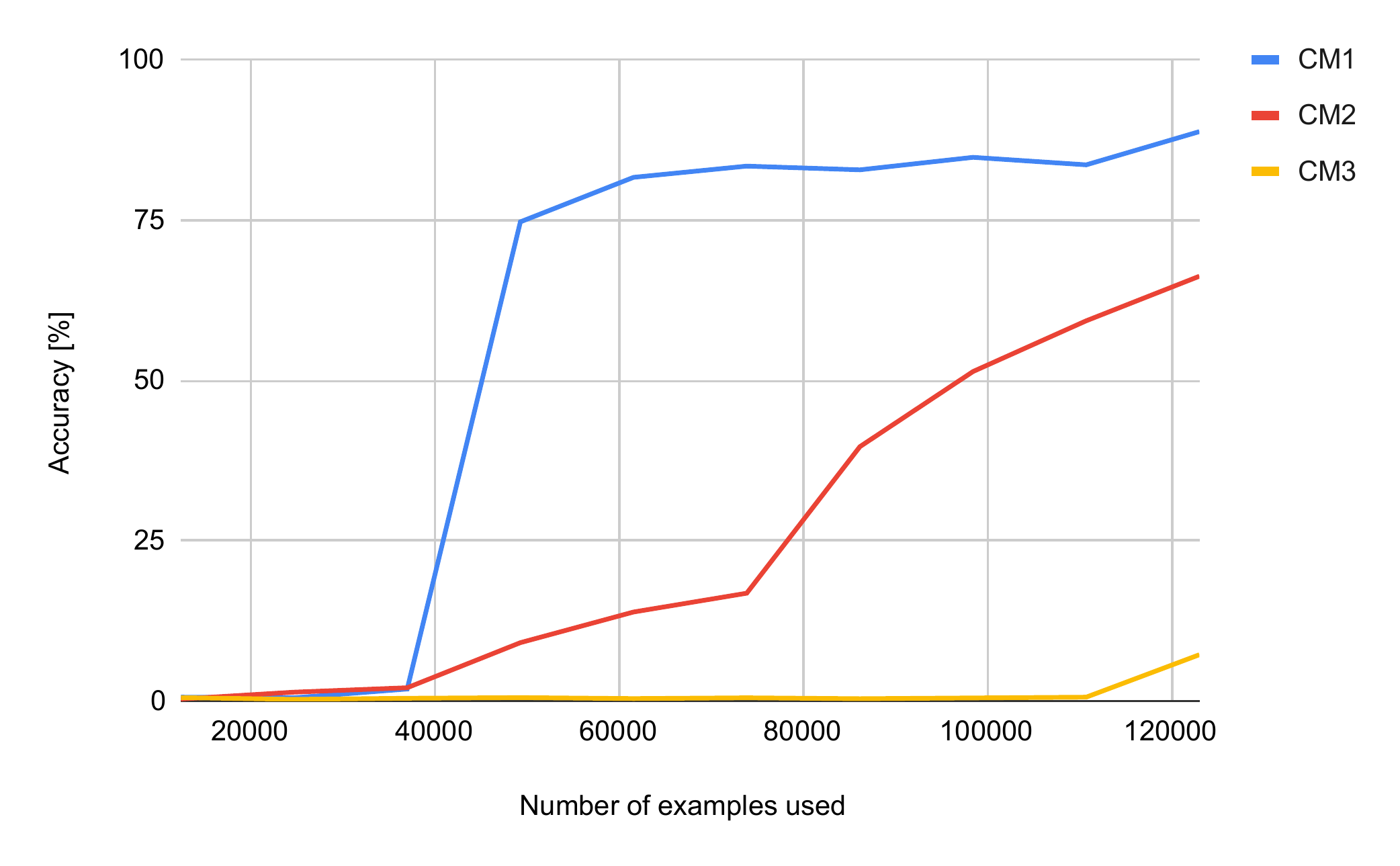}%
        \vspace{-1.2cm}
    }\hfill
    \subfigure[MeanRank]{%
        \label{fig:parts_of_ds_2rounds_k_0_mean_rank}%
           \vspace{-.4cm}
        \includegraphics[width=0.47\textwidth]{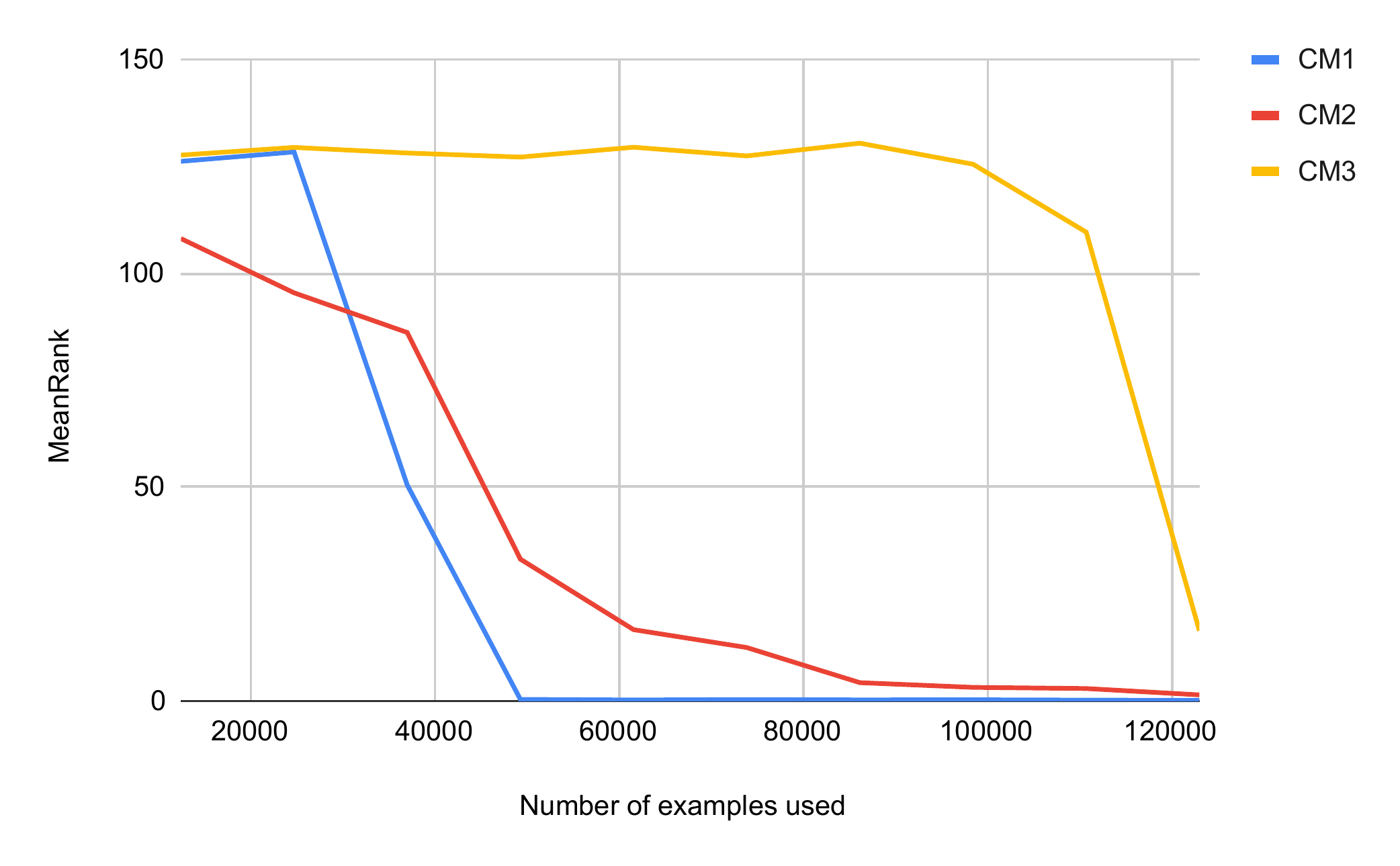}%
          \vspace{-1.2cm}
    }
     
    \caption{ECC $k_0$ accuracy and MeanRank while using only a part of the dataset. Results evaluated on the test split.}
    \label{fig:my_label}
\end{figure}

We found out, as visible in
\Cref{fig:parts_of_ds_2rounds_k_0_acc} and \Cref{fig:parts_of_ds_2rounds_k_0_mean_rank}, that \modelname{} is most likely bounded by the lack of data since the accuracy keeps rising as the amount of training examples increases.
In particular, for CM3, \modelname{} starts to generalize only when at least 110,000 examples are used, suggesting that increasing the dataset size would most likely lead to significant accuracy gain.
However, the CM3 dataset already requires 3.7TB of storage, so we decided against increasing the dataset size further, as \modelname{} is already able to successfully attack CM3, and increased accuracy does not bring significant additional benefits.

\subsection{ECDSA attack}
\label{sec:ecdsa_attack}
Due to the presence of highly-secure countermeasures, \modelname{} is not able to recover all bytes of the multiplier at once with sufficient confidence.
Thus, a single-trace attack similar to \cite{EPRINT:WCPB21} seems out of reach. This is expected, as custom attack models have an edge over generalized models in this regard.
However, a multi-trace attack is feasible, as we shall demonstrate. Specifically, the partial leakage we obtained can be combined with a lattice attack to recover the key from about 8000 traces, as demonstrated by \cite{moghimi2020tpm,roche2021side}.

We apply a standard lattice attack~\cite{howgrave2001lattice,nguyen2002insecurity} to leverage the partial leakage from ECC scalar multiplication protected by either CM1, CM2 (even blackbox), or CM3 and recover the private key from ECDSA (\cite{johnson2001elliptic}).
To simulate a realistic attack, we take our holdout split (captured on a different physical chip than the training data), treat it as the nonce multiplication of ECSDA $k \times G$ explained in \Cref{back:ecdsa}, and predict the most significant byte.
One caveat is that our multipliers in the holdout split are chosen to test the deep learning model so that each byte is chosen independently and uniformly at random between 0 and 255.
Since for ECDSA we require the nonce $k$ to be $1 \leq k < n$, where $n$ is the size of the elliptic curve, we exclude all measurements that have the multiplier outside of this range, after which we are left with roughly 7,800 examples (depending on the dataset – CM1, CM2, CM3).
We try to closely simulate an attacker with a profiling device (i.e., the chip the attacker uses to collect training and validation splits) and traces from the device under attack.
The attacker is free to leverage the model's outputs as they wish, as long as they are able to recover the secret key in reasonable time.
The way of using prediction confidence described below is specific to our attack configuration (model and dataset), and we do not claim that it directly transfers to other attack configurations.

The lattice attack requires all predictions to be correct for it to work, or as shown by~\cite{dall2018cachequote}, it can tolerate a very small number of noisy signatures.
After trial and error with the predictions, our intuition is that if we turn byte predictions into predictions of the four most significant bits (MSBs), we will achieve the highest accuracy.
To compute the probability of the 4 MSB of the nonce for a given signature, we sum the probabilities of the 16 possible LSB values with the same 4 MSBs.
\Cref{tab:ecdsa_4msb_results} shows the prediction accuracy of the four most significant bits.
Here, we see much higher accuracy.

\begin{table}[htb]
  \centering\tiny
   \fullVersionOnly{\scriptsize}
    \begin{tabular}{p{4cm}rr}
         \textbf{Experiment} & \textbf{Accuracy [\%]} & \textbf{MeanRank} \\
         \hline
         CM1 & 94.83 & 0.06 \\
         CM2 & 96.39 & 0.07 \\
         CM2 black-box & 85.75 & 0.20 \\
         CM3 & 71.86 & 0.87 
    \end{tabular}
    \caption{Results of predicting the 4 most significant bits (on holdout).}
    
    \label{tab:ecdsa_4msb_results}
\end{table}%

A lattice attack using the four MSBs of the nonce requires 80 signatures~\cite{jancar2020minerva,moghimi2020tpm} predicted by the model.
Still, random sampling of 80 signatures will have a high chance of having erroneous signatures, especially when the accuracy is still under 90\%.
However, we notice that in cases where the correct key 4 MSBs are detected with high accuracy, the prediction value has a higher  \emph{confidence} (highest probability - second highest probability).
We use the confidence of predictions as weights when randomly sampling (using the parameter {\tt weights} of {\tt random.choices} in Python).
This approach retrieves the secret key after several retries for CM1 and CM2.

For the case of CM3, we employ the following heuristics to succeed in the attack.
First, we give more weight to samples with higher confidence; we achieve this by using confidence to the power of eight as weights when sampling for the subset used in the lattice attack.
The constant eight is chosen by profiling the attack on the validation set (also roughly 8,200 examples).
Second, we discard samples with too high confidence (more than 0.25, chosen by trial and error) when predicting the byte value,  discarding roughly 10\% of examples.
This excludes all (and thus also the wrong) predictions that are too confident and would occur in most random samples.

We use the lattice construction of \cite{benger2014ooh} and \cite{nguyen2002insecurity} with the \cite{sagemath} implementation of the BKZ algorithm \cite{schnorr1994lattice}.
The attacks take a few minutes.
The longest is the black box attack on ECDSA using CM2, which takes roughly 15 minutes on a desktop computer with AMD Ryzen 9 CPU.
\fullVersionOnly{This is due to several retries needed to get all 80 samples correct.}

\callout{Additional evaluation on a public ECC dataset.}
\label{sec:reassure_ecc}
To better compare to prior work, we evaluate \modelname{} against the REASSURE ECC dataset introduced by \cite{chmielewski2020reassure}. This datasets consists of roughly 6,000 unprotected traces subdivided into 255 sub-traces that are aligned to expose the corresponding cswap bit of the Montgomery ladder. Each of the sub-traces consists of 5,500 points.\fullVersionOnly{

}
The model described in \cite{SAC:NCOS16} predicts the cswap bits with $99.6\%$ accuracy and therefore recovers the whole key from a single trace. \modelname{} achieves comparable accuracy $99.57\%$ \emph{without the need for hypertuning}, showcasing once again its ability to generalize across datasets and use-cases.

\callout{Evaluating prior-art models on our datasets.}
\label{sec:other_models_on_our_ecc_datasets}
Finally, to evaluate the difference in generalization capability between \modelname{} and previous work, we train the LSTM model proposed in~\cite{lu2021pay}
on ECC CM0, targeting $k_0$. This model  
achieves $91.4\%$ accuracy but fails at attacking CM1. This  demonstrates its limitations in generalizing past simple constant-time defenses to strongly protected implementations.
On the other hand it proves our point that a single model can be used for multiple algorithms (\cite{lu2021pay} were targetting AES).

We also train a CNN model similar to \cite{EPRINT:WCPB21} to evaluate convolution networks generalization potential. We had to halve the number of filters and lower the batch size to 32 to be able to run it on our longer ECC traces. This model achieved $100\%$ accuracy on CM0 $k_0$ but, similarly to \cite{lu2021pay}, failed to show significant leakage on CM1 $km_0$ (mean rank $118$). This highlights a similar limited ability to generalize against strong defenses.
\fullVersionOnly{
We note that those finding are consistent with our extensive internal experiments before landing on \modelname{}, during which we were not able to get even state-of-the-art convolutional architecture such as ConvNeXt \cite{liu2022convnet} to achieve results as good as \modelname{}'s.
}

\section{Generalizing \modelname{} to AES} \label{sec:aes}

In this section, we show that \modelname{} generalizes across cryptographic algorithms by demonstrating its effectiveness in attacking an AES software-protected implementation, namely the publicly available ASCADv2 dataset~\cite{masure2023side}, in an end-to-end manner without trace processing.
\fullVersionOnly{Additionally we evaluate \modelname{} on two other publicly available AES datasets namely ASCADv1 \cite{emmanuel2018study} and CHES 2023 challenge to assess its ability to generalize to various AES implementations and compare its performances to previous specialized approaches.}

\callout{AES attack evaluation.}
Leakage from a single trace is often not enough to uncover the secret key in AES.
In this case, the attacker may combine information from multiple traces with the same key but variable plaintext.
Predictions of key byte values are then combined over a set of attack traces to a maximum likelihood score vector (their logarithms are summed).
The index in this vector with the largest value is the predicted value of the key byte.
More generally, \emph{Guessing Entropy} (GE) \cite{EPRINT:StaMalYun06} is the average number of entries larger than the one corresponding to the correct value (also mean rank of the correct value in the score vector).
When the GE of all key bytes is less than one we call the attack successful.  When a sensitive value $s$ (e.g., value of a byte of the S-BOX input) is split into two shares, bytes $x, y$, we may target $x$ and $y$ separately and then from their probabilities we compute $P[s = b] = \sum_{i = \texttt{0x00}}^{\texttt{0xFF}} P[x = b \oplus i] P[y = i]$ for any byte value $b \in \{ \texttt{0x00}, \ldots, \texttt{0xFF} \}$.
Analogous formulas for other types of masking (shuffling and affine masking) are derived by \cite{masure2023side}.

\subsection{ASCADv2 dataset}
\label{sec:ascad_dataset}
The ASCADv2 dataset~\cite{masure2023side} comprises 800,000 power traces collected from a Cortex M4 microcontroller manufactured by ST Microelectronics (STM32F303RCT7) while it was performing AES-128 encryptions.
The firmware implements affine masking and shuffling to protect the AES encryption computation from side-channel attacks.
More details about the dataset can be found in~\cite{benadjila2020hardened}.

\callout{Attack Scenario.}
We replicate the ``First Threat Scenario'' described in~\cite{masure2023side}.
We target the following equation from the AES S-BOX masked operation~\cite{masure2023side}:
$
    c[i] = r_m \times Sbox[pt[p[i]] \oplus k[p[i]]] \oplus r_{out}
$ ,
where $\times$ and $\oplus$ stand for multiplication and addition in the Rijndael finite field~\cite{daemen2001reijndael},  $Sbox$ is the AES S-BOX,   $r_{m}$ and $r_{out}$ are affine mask bytes, 
  $p[i]$ is the permutation index,
  $pt[i]$ is the byte of the plaintext, and
  $k[i]$ is the byte of the AES round key.
\fullVersionOnly{

}
Critically, unlike previous work~\cite{masure2023side}, we perform an attack by using all $1,000,000$ points of each trace without preprocessing, instead of using an SNR (signal-to-noise ratio analysis) analysis to use only $15,000$ points (1.5\%) out of the total trace.
Additionally, we do not modify the \modelname{} architecture to perform the attack and solely rely on hyperparameter tuning to adjust the model hyperparameters.

\begin{table}[htb]
   \centering\tiny
   \fullVersionOnly{\scriptsize}
    \begin{tabular}{lrrrrr}

        \textbf{Attack point } & \textbf{Accuracy  [\%] } & \textbf{MeanRank}& \textbf{MaxRank} & \textbf{Acc~\cite{masure2023side}} & \textbf{MeanRank~\cite{masure2023side}} \\
      
        \hline
        $r_m$ & \best{100.00} & 0 & 0 & 99.2 & -- \\
        $r_{out}$ & 18.25 & 4.78 & 65  & \best{21.1} & -- \\
        $c[i]$ (average) & 1.18 & 80.65 & 255 & \best{1.6} & \best{80} \\
        $p[i]$ (average) & \best{95.07} & 0.055 & 4.44 & 88.9 & -- \\
    \end{tabular}
    
    \caption{ASCADv2 results ($c[i]$ is the best out of 7 runs) compared with results of~\cite{masure2023side}, measured on holdout dataset with 80,000 examples. For the sake of brevity we average results when there are multiple indexes.}
    \label{tab:ascadv2_best_results}
\end{table}

\callout{AES attack evaluation.}
\label{sec:ascad_attack}
For $c[i]$, we report the best of 7 runs (due to the influence of initialization weights, see \Cref{sec:initialization_dependence}).
That is, we train a model seven times, and for each attack point, we pick the model with the highest accuracy on the validation set.
We then use that to evaluate the performance on the holdout split.
Our model targets the intermediate value $c[i]$, mask bytes $r_m, r_{out}$, and the permutation $p[i]$ (for $i = 0, \ldots, 15$).
\Cref{tab:ascadv2_best_results} shows a comparison of ML metrics of \modelname{} and the results of~\cite{masure2023side}.
We then estimate GE to also compare attack results from the acquired intermediate values.
Following \cite{masure2023side} we simulate a fixed key (all zero) by replacing $pt$ by its XOR with the corresponding random key.
Since $r_m$ has very high accuracy we use it directly instead of the whole probability distribution.
We sample 10,000 times to estimate the GE and need roughly 80 traces to have GE of all key bytes under 1, comparable with the 60 traces needed by~\cite{masure2023side} (but use heavy preprocessing).

\subsection{ASCADv1 variable key}
\label{sec:ascad_v1}
ASCADv1 the precursor of ASCADv2 is comprised of two electromagnetic emission datasets captured of an ATmega8515 micro-controller running a masked AES implementation. Following~\cite{egger2022second} we target the dataset with variable key and compare to the SOTA attacks \cite{lu2021pay} and \cite{hajra2024estranet}.
For this attack we hypertune the patch size (possible values $[100, 200, 400, 625, 1000, 2000]$) and merge filter 1 (possible values $[0, 4, 8, 16, 32, 64]$) 49 experiments in two days, winner merge filter = 0 and patch size 625.
Each tuning run comprise 50 epochs of 500 steps and use a batch size of 256.
The learning rate is set to $0.0005$ and we use the full trace length.
This configuration was fully trained for 500 epochs and achieves a $95.94\%$ accuracy on the third byte of S-BOX input which is the standard target as the first two bytes in the dataset are unprotected due to the mask being always zero.
\modelname{} significantly outperforms both \cite{lu2021pay} and \cite{hajra2024estranet} neither of which achieves a single trace attack.

\subsection{CHES 2023 SMAesH challenge}
\label{sec:ches_23}
The challenge\footnote{\url{https://smaesh-challenge.simple-crypto.org/}} consists of two protected AES datasets collected from an A7 (Artix-7 FPGA) and S6 (Spartan-6 FPGA) target boards.
The primary goal is to mount a key recovering attack by reducing the key rank below $2^{68}$, with a secondary objective to minimize the number of traces used. To make a fair comparison,  we measure our key rank reduction while using the same number of traces as the best submission (A7: 290k traces using \cite{AC:VeyGerSta14}, S6: 901k traces using \cite{EPRINT:MCLS22}). Within these constraints, \modelname{} correctly identifies leakage in both cases (achieving its primary goal) but does not fully recover the key without customization.
\fullVersionOnly{
We note that \modelname{} prefers generalizability over attack-specific customization, and as such it is not designed to emphasize the minimization of the number of attack traces.
}

\callout{A7 dataset}. As indicated, we target bytes $M[i] = \texttt{msk\_plaintext}[i] \oplus \texttt{msk\_key}[i]$ (since $M[i] \oplus M[i+16] = \texttt{key}[i] \oplus \texttt{plaintext}[i]$).
Accumulating predictions of $M[i] \oplus M[i+16]$ from predictions of $M[i], M[i+16]$ sometimes fails and requires re-training.
It is an open question if \modelname{} training would be more stable using the technique of \cite{EPRINT:MCLS22}.
GE at 290k traces is consistently between 20 and 50 which is too high for an attack (key rank estimate $2^{70}$ to $2^{90}$).

\callout{S6 dataset}. As indicated in the competition report \cite{ches23challengeslides},  we target bytes $S[i] = \texttt{key}[i] \oplus \texttt{plaintext}[i], i \in \{ 1, 6, 11, 12 \}$. Doing so, we fully uncover these four key bytes.
Targeting $S[1] \oplus S[5]$ while fully knowing $S[1]$ reduces our guessing entropy of $\texttt{key}[5]$ to 4.4.
Other attack points $S[i] \oplus S[i + 4 \mod 16]$ would likely benefit from re-training.
The full key rank we achieve is estimated to be $2^{90.13}$.

\section{Conclusion} \label{sec:conclusion}

In this paper we presented \modelname{}, the first deep-learning architecture that can perform power side-channel analysis against multiple protected cryptographic algorithms, namely ECC and AES. 
We  demonstrate \modelname{}'s ability to generalize by successfully attacking several highly protected ECC implementations without changing the model architecture, and verify its effectiveness on multiple devices.
\fullVersionOnly{To enable reproducibility of these results, we open-source our models and datasets.}
This research moves us one step closer to a fully-automated side-channel attacks and leakage detection system that could be used as part of a hardware product release process and to test new countermeasures.
Our results  suggest that some advanced countermeasures that are  considered sufficient to thwart side-channels attacks can be defeated.
Accordingly, there is a pressing need to devise new countermeasures that are resilient to ML attacks.

\bibliographystyle{alpha}
\bibliography{abbrev3,crypto,references}

\end{document}